\documentclass[traditabstract]{aa} 
\usepackage{graphicx}
\usepackage{txfonts}
\usepackage{longtable,lscape}
\usepackage{natbib}
\bibpunct{(}{)}{;}{a}{}{,}
\newfont{\gwpfont}{cmssq8 scaled 1000}
\newcommand{\rexcess}{{\gwpfont REXCESS}}
\def \xmm {\hbox{\it XMM-Newton}}
\def \chandra {\hbox{\it Chandra}}

\begin{document}
   \title{The MCXC: a Meta-Catalogue of X-ray detected Clusters of galaxies}

\author{R. Piffaretti\inst{1}, M. Arnaud\inst{1}, G.W. Pratt\inst{1}, E. Pointecouteau\inst{2} and J.-B. Melin\inst{3}
}
\offprints{R. Piffaretti, \email{rocco.piffaretti@cea.fr}}
\authorrunning{R. Piffaretti et al.}
\titlerunning{The MCXC}
 \institute{
 $^1$ Laboratoire AIM, IRFU/Service d'Astrophysique - CEA/DSM - CNRS - Universit\'{e} Paris Diderot, B\^{a}t. 709, CEA-Saclay, F-91191 Gif-sur-Yvette Cedex, France \\ 
 $^2$ Universit\'{e} de Toulouse, CNRS, CESR, 9av. du colonel Roche, BP 44346, 31028 Toulouse Cedex 04, France \\
$^3$ DSM/Irfu/SPP, CEA/Saclay, F-91191 Gif-sur-Yvette Cedex, France\\
}

   \date{Received ; accepted }
 
\abstract{

We present the compilation and properties of a Meta-Catalogue of X-ray detected Clusters of galaxies, the MCXC. This very large catalogue is based on publicly available ROSAT All Sky Survey-based (NORAS, REFLEX, BCS, SGP, NEP, MACS, and CIZA) and serendipitous (160SD, 400SD, SHARC, WARPS, and EMSS) cluster catalogues. Data have been systematically homogenised to an overdensity of 500, and duplicate entries originating from overlaps between the survey areas of the individual input catalogues are carefully handled. The MCXC comprises 1743 clusters with virtually no duplicate entries. For each cluster the MCXC provides: three identifiers, a redshift, coordinates, membership of original catalogue, and standardised $0.1-2.4$ keV band luminosity $L_{\rm 500}$, total mass $M_{\rm 500}$, and radius $R_{\rm 500}$. The meta-catalogue additionally furnishes information on overlaps between the input catalogues and the luminosity ratios when measurements from different surveys are available, and also gives notes on individual objects. The MCXC is available in electronic format for maximum usefulness in X-ray, SZ, and multi-wavelength studies.
}

\keywords{Catalogs, Cosmology: observations, Cosmology: large-scale structure of Universe, Galaxies: cluster: general - X-rays: galaxies: clusters}

   \maketitle
%
%
\section{Introduction}
\label{intro.sec}

Clusters of galaxies provide cosmological constraints through the number density and evolution of objects, through the power spectrum of their three-dimensional distribution, and through their baryon fraction and its evolution. Moreover, the physical properties of clusters provide a test of the structure formation scenario, giving vital information both for understanding the gravitational collapse of the dark matter and for the evolution of baryons in the dark matter potential \citep[see][for a review]{voitrev2005}.

X-ray observations are ideal for these studies as the density squared dependence of the X-ray emission means that clusters can efficiently be found over a wide redshift range. Cluster sources were evident in the first all-sky X-ray survey with {\it Uhuru}, and further objects were found by HEAO-1 and {\it Ariel-V}; subsequent follow-up observations with {\it Einstein} and {\it EXOSAT} allowed more accurate characterisation of their physical properties \citep[see][for a review]{rosati2002}.

In this context, the ROSAT satellite has played a central role. The 1990-1991 ROSAT All-Sky Survey \citep[RASS,][]{rass} and later deep pointed observations have led to the discovery of hundreds of clusters. Subsequent follow-up observations, in particular those conducted with the current generation of X-ray satellites \xmm, \chandra\ and {\it Suzaku}, have provided statistical samples for cosmological studies \citep[e.g.,][]{vik2009, mantz2} and detailed information on the structural properties of the cluster population \citep[e.g.,][]{vikh06,pap10,app2009}. Other observations have allowed in-depth study of the hierarchical assembly process through merging \citep[e.g.,][]{mv07} and the physical mechanisms associated with feedback and its impact on structure formation \citep[e.g.,][and references therein]{mcnamara2007}. However, while several \xmm\ and \chandra\ X-ray surveys are ongoing \citep[e.g.,][]{xcs2001,barkhouse2006,pacaud2007,xdcp2008}\footnote{See {\tt http://cxc.harvard.edu/xraysurveys/surveys.html} for a complete list of ongoing \xmm\ and \chandra\ surveys.}, the associated cluster catalogues are either not yet published or only partially available. 

Outside of the X-ray domain, the redshift-independent thermal Sunyaev-Zel'dovich effect \cite[][hereafter SZ]{sz1972} is emerging as an efficient way to detect distant, massive clusters that fall below the flux limits of X-ray surveys. Several SZ surveys, including the South Pole Telescope \citep[SPT,][]{carlstrom2009} survey, the Atacama Cosmology Telescope \citep[ACT,][]{fowler2007}, and Planck \citep{tauber2010}, are actively ongoing and have started providing the first SZ-selected cluster catalogues \citep[e.g.,][]{vanderlinde2010,menanteau2010}. X-ray observations of SZ clusters are important in many respects. The X-ray properties allow a better characterisation of the SZ signal \citep[e.g.,][]{melin2006,andersson2010} and yield the calibration of the scaling relations needed for cosmological studies with SZ-selected cluster samples \citep[e.g.,][]{majumdar2003}. In addition, X-ray observations allow testing of the selection function of SZ surveys and verification of new SZ cluster candidates \citep[e.g.,][]{suh10}. Moreover, they are essential for statistical analyses of the SZ data \citep[e.g.,][and references therein]{melin10,komatsu2010}.

Cosmological tests that rely on knowledge of the evolution of the mass function or baryon fraction require an estimate of the cluster mass. Surveys provide only an observable (typically luminosity, temperature or SZ $y$-parameter) that is then linked to the cluster mass via scaling relations. While simultaneous constraints on cosmological parameters and scaling relations have recently been derived \citep{mantz1}, the mass proxy relations are typically separately calibrated using deep observations of well-understood and if possible representative samples \citep[e.g.,][]{arnaud2007,maughan2007,vik2009,pratt2009}. Although their redshift evolution is at present poorly known, a consensus on the type of scaling relations to be calibrated and their precise definition has been reached. For example, the bias introduced by cool core clusters in luminosity and temperature measurements is taken into account, low scatter mass proxies such as $Y_{\rm X}$ \citep{kravtsov2006} or the gas mass $M_{\rm gas}$ are widely used, and all quantities are measured up to a standard characteristic radius $R_{\rm 500}$, the radius within which the mean over-density of the cluster is 500 times the critical density at the clusterÕs redshift. Substantial progress has also been made in understanding the systematics affecting X-ray mass estimates via simulations \citep[e.g.,][]{rasia06,nagai07,pv2008} and via combination with gravitational lensing \citep[e.g.,][]{mahdavi2008,zhang2010,meneghetti2010}. 

ROSAT-derived catalogues still play a major role in providing targets for deeper observation with the current generation of X-ray 
 instruments \citep[e.g.,][]{bohrexcess,vik2009}, and for identification of existing clusters in new surveys in other wavelength bands \citep[e.g.,][]{popesso2004}. These catalogues have been derived from a number of surveys based on RASS data or ROSAT pointed observations (see Sect. \ref{catalogues.sec}). Despite the fact that these catalogues are public, no attempt has yet been made to merge them and to homogenise the data contained within. The main reasons for this are their large sizes and the fact that different catalogues provide different types of information.

Given the current status of X-ray cluster catalogues, and the relevance of scaling relations for SZ surveys and X-ray studies in general, we collected data from all the major public X-ray survey catalogues and homogenised the information to provide the community with a meta-catalogue of X-ray detected clusters of galaxies (hereafter the MCXC). The basic characteristics of the MCXC are the large number of clusters in the catalogue (1743 unique systems), a uniform format for all provided quantities, careful control of duplicate entries originating from overlaps between the input catalogues, and homogeneously estimated $0.1-2.4$ keV band luminosities $L_{\rm 500}$ and total masses $M_{\rm 500}$. In order to be easily manipulated, the MCXC is provided in electronic format. The final catalogue gives a first overview of the published, publicly-available X-ray survey selected cluster population.

The paper is organised as follows. In Sect. \ref{catalogues.sec} we describe the basic properties of the catalogues used to construct the MCXC. In Sect. \ref{data.sec} we explain how the information is homogenised and detail the quantities provided by the MCXC. The handling of duplicate entries is presented in Sect. \ref {overlap.sec} and in Sect. \ref{discussion.sec} we discuss various aspects of the final catalogue. In Sect. \ref{conclusions.sec} we summarize our results and present our conclusions.

As a cosmological model we adopt a $\Lambda$CDM cosmology with $H_{0} = 70$~ km/s/Mpc, $\Omega_{\rm M} = 0.3$ and $\Omega_\Lambda= 0.7$ throughout the paper. The quantity $h(z)$ is the ratio of the Hubble constant at redshift $z$ to its present value, $H_{0}$, i.e., $h(z)^2=\Omega_\mathrm{m} (1+z)^3 + \Omega_\mathrm{\Lambda}$.

\section{Input X-ray catalogues}
\label{catalogues.sec}

In the following we describe the input catalogues used to construct the MCXC. We recall the basic characteristics of the X-ray surveys used to construct each catalogue and how the X-ray quantities adopted in our work are measured. We discuss only X-ray information essential to the MCXC and focus on the quantities that allow us to compute the luminosities, $L_{\rm 500}$. For more details on the individual surveys, and in particular the associated optical observations/followup, we refer the reader to the cited papers and references therein.

Generally speaking, two types of X-ray survey can be distinguished : contiguous area surveys, which use data from the ROSAT All-Sky Survey \citep[RASS,][]{rass}, and serendipitous cluster surveys, which are based on data from deeper pointed X-ray observations. In the following, we therefore distinguish between RASS-based and serendipitous catalogues. In addition to handling duplicate entries and removing particular clusters as discussed below, we exclude clusters with non measured redshifts or luminosities. Table \ref{table:nrobj} summarises the contributions of the various input catalogues to the MCXC.

The bulk of the X-ray data used to construct the MCXC are derived from ROSAT observations. Exceptions are EMSS and some physical quantities for MACS, as described in more detail below. Future work will include as-yet unpublished catalogues such as RDCS \citep[the ROSAT Deep Cluster Survey,][]{rdcs1998}, XCS \citep[the XMM Cluster Survey,][]{xcs2001}, XDCP \citep[the XMM-Newton Distant Cluster Project,][]{xdcp2008}, and the complete MACS catalogue \citep{macs2001}.

\subsection{RASS-based catalogues}

We compiled data from nine RASS-based contiguous area surveys, as described below.

\subsubsection{REFLEX and NORAS}

{\bf REFLEX} (ROSAT-ESO Flux Limited X-ray Galaxy Cluster Survey, \citealt{reflex}) is based on RASS data for a survey area covering the southern sky up to a declination $\delta= 2.5 \, {\rm deg}$  with the galactic plane ($\mid b \mid \le 20 \, {\rm deg}$) and the regions of the Magellanic clouds excluded. The total survey area is 13924 deg$^2$ and the survey is flux-limited ($0.1-2.4$ keV band flux $ \ge 3 \times 10^{-12} \, {\rm erg} \, {\rm s}^{-1} \, {\rm cm}^{-2} $). 

{\bf NORAS} (Northern ROSAT All-Sky galaxy cluster survey \citealt{noras}) is also based on RASS data excluding the same region around the galactic plane, but covers the northern sky.  This survey catalogue is not flux-limited and selection is based on minimum count rate ($0.06 \, {\rm cts/s}$ in the $0.1-2.4$ keV band) and a source extent likelihood.

The data analysis and catalogue production for both NORAS and REFLEX are performed by essentially the same authors and, although REFLEX has been more extensively studied and characterised than NORAS, the information provided is extremely similar. A growth curve analysis is adopted to determine source fluxes (the typical flux measurement accuracy is $10-20 \%$) and luminosities. The REFLEX catalogue provides aperture luminosities $L_{\rm ap}$ as well as total luminosities. The latter are computed by estimating the missing flux outside the detection aperture by assuming a $\beta$-model with $\beta=2/3$, a core radius $r_{\rm c}$ which scales with mass, and a cluster extent of $12 \times r_{\rm c}$. For the NORAS clusters a similar procedure is performed, but the resulting total luminosities are not reported. Therefore the NORAS catalogue provides only aperture luminosities.  

For both catalogues data \citep{norasC, reflexC} are retrieved from VizieR\footnote{\tt http://VizieR.u-strasbg.fr/viz-bin/VizieR}. Because of the homogeneity of these two catalogues we merge them into a single NORAS/REFLEX catalogue. The names {\tt NORAS} and {\tt REFLEX} are kept as sub-catalogue labels. Because of the overlap of the NORAS and REFLEX survey areas, there are ten duplicate entries. For these ten clusters the information provided by NORAS and REFLEX is almost identical and we exclude, for each of the duplicates, the cluster with the larger flux uncertainty.

Since the number of clusters in the combined NORAS/REFLEX catalogue is large (889 objects, see Table \ref{table:nrobj}) and because the information provided by the authors is homogeneous and detailed, it is the cornerstone of the MCXC.

\subsubsection{ROSAT BCS and eBCS}

The {\bf ROSAT BCS} (The ROSAT Brightest Cluster Sample, \citealt{bcs})  comprises the brighter sources of the NORAS survey. We use data for the 90 per cent complete BCS, a flux-limited sample ($0.1-2.4$ keV band flux $ \ge 4.4 \times 10^{-12} \, {\rm erg} \, {\rm s}^{-1} \, {\rm cm}^{-2} $) of $z \le 0.3$ clusters. 

The {\bf eBCS} (The extended ROSAT Brightest Cluster Sample, \citealt{ebcs}) is the low flux extension of the BCS ($0.1-2.4$ keV band flux $ \ge 2.8 \times 10^{-12} \, {\rm erg} \, {\rm s}^{-1} \, {\rm cm}^{-2} $).

The type of information provided is the same for both samples. In both cases detection and cluster emission characterisation are based upon the Voronoi tesselation and percolation (VTP) algorithm. The emission outside the detection region is computed by correcting the detected count rate. For clusters this is undertaken by assuming a $\beta$-model profile with $\beta=2/3$ and a core radius estimated from the source profile, taking into account the telescope PSF. The resulting total luminosities, the corrected and uncorrected count rates, and the VTP aperture radius are provided. This implies that the luminosity within the VTP aperture radius can be computed for all the clusters in the sample.

Data \citep{bcsC, ebcsC} are retrieved from VizieR and merged into a single BCS catalogue where the names {\tt BCS} and {\tt eBCS} are kept as sub-catalogue labels (see Table \ref{table:nrobj}). Note that there is only one cluster, A1758a, that is listed in both BCS and eBCS. The two luminosities are almost identical and we remove it from the BCS sub-catalogue. In addition, for the Virgo cluster we adopt the luminosity estimate of \cite{boh1994}.

\subsubsection{SGP}

The {\bf SGP} (A Catalog of Clusters of Galaxies in a Region of 1 Steradian around the South Galactic Pole, \citealt{SGP}) covers a region of 1.013 sr centered on the south Galactic pole and is based on the same X-ray source detection and characterisation procedures as REFLEX. The lowest detected flux is  $ 1.5 \times 10^{-12} \, {\rm erg} \, {\rm s}^{-1} \, {\rm cm}^{-2} $ in the $0.1-2.4$ keV band, and a complete sub-sample can be obtained by imposing a flux limit of $ 3 \times 10^{-12} \, {\rm erg} \, {\rm s}^{-1} \, {\rm cm}^{-2} $. 

Luminosities are computed within a cutoff radius provided by the growth curve analysis. Since the cutoff radius is not given in the catalogue, we treat the quoted luminosity as the total luminosity. 

Data for the entire non-flux-limited, SGP sample \citep{SGPC, SGPerr} were retrieved from VizieR.

\subsubsection{NEP}

The {\bf NEP} (The ROSAT North Ecliptic Pole survey, \citealt{nep2006}) surrounds the north ecliptic pole in a survey area of 80.6 deg$^2$, and has the deepest exposure in the northern RASS (exposure times from 2\,000 to over 40\,000 s). Source detection is based on \cite{rass} and the selection is performed by adopting thresholds for the source extent likelihood and signal-to-noise ratio. 

The quoted total luminosities are computed from size corrected fluxes. The latter are computed from detected fluxes within apertures of radius 5' (6.5' for RXJ1834.1+7057) by assuming a PSF-corrected $\beta$-profile with $\beta=2/3$ and a fixed core radius of 180 kpc. The profile is integrated up to $R_{\rm 200}$, which is estimated from the size-temperature relation of \cite{evrard1996}. Size correction factors are provided so that aperture luminosities can be computed.

Data for the whole flux-limited sample ($0.5-2$ keV band flux $ \ge 2. \times 10^{-14} \, {\rm erg} \, {\rm s}^{-1} \, {\rm cm}^{-2} $, \citealt{nep2006}) were retrieved from VizieR, and only sources identified as clusters were selected (see Table \ref{table:nrobj}).

\subsubsection{MACS}

The {\bf MACS} (Massive Cluster Survey, \citealt{macs2001}) is based on the ROSAT Bright Source Catalogue with the aim of increasing the number of known very luminous, $z \ge 0.3$ clusters. A MACS catalogue has not yet been published in its entirety and we therefore collected data from different publications as detailed below. Notice that the data reported in these publications are based on {\it Chandra} follow-up observations and that these publications yield all publicly-available MACS clusters with coordinates, redshifts, and luminosities (i.e., the minimal set of quantities required for the MCXC). 

Properties of a complete subsample of $z > 0.5$ MACS clusters (the {\tt MACS\_DIST} sub-catalogue, twelve objects) are listed in \cite{ebeling2007}. A further complete subsample of bright objects in the $0.3 < z < 0.5$ redshift range (the {\tt MACS\_BRIGHT} sub-catalogue, 34 clusters) are given in \cite{macs2010}. For these sources, the luminosities within $R_{\rm 200}$ are listed. For 32 of the 34 {\tt MACS\_BRIGHT} clusters \cite{mantz2} give additional properties such as $L_{\rm 500}$, $M_{\rm 500}$, etc\footnote{MACSJ0358.8-2955 and MACSJ2311.5+0338 are not studied in \citet{mantz2}.}. This information is also merged into the {\tt MACS\_BRIGHT} sub-catalogue.

Further MACS clusters are analysed in \citet[][the {\tt MACS\_MJFV} sub-catalogue, 23 objects]{maughan2008}, who provide very complete information on the physical properties of these objects. Of the {\tt MACS\_MJFV} sample there are six clusters in common with the {\tt MACS\_DIST} sub-catalogue and twelve clusters in common with the {\tt MACS\_BRIGHT} sub-catalogue. We construct a unique MACS catalogue by merging the three sub-catalogues and keeping only measurements given by \citet{maughan2008} for the eighteen duplicate clusters (see Table \ref{table:nrobj}). Apart from the six {\tt MACS\_DIST} luminosity measurements in \citet{ebeling2007}, the luminosities $L_{\rm 500}$ are directly available for all MACS clusters.

\subsubsection{CIZA}

The {\bf CIZA} (Clusters in the Zone of Avoidance, \citealt{ciza1} and \citealt{ciza2}, respectively {\bf CIZAI} and {\bf CIZAII}) catalogues are based on the ROSAT Bright Source Catalogue and focus on the region around the galactic plane ($\mid b \mid \le 20 \, {\rm deg}$). Candidate selection is based on limits on the detected fluxes and spectral hardness ratios. CIZAI comprises the X-rayÐbrightest objects (flux $ \ge 5. \times 10^{-12} \, {\rm erg} \, {\rm s}^{-1} \, {\rm cm}^{-2} $), while CIZAII is its low-flux extension (flux $ \ge 3. \times 10^{-12} \, {\rm erg} \, {\rm s}^{-1} \, {\rm cm}^{-2} $). 

Quoted luminosities are computed from raw RASS data using very large apertures and can be therefore safely interpreted as total luminosities.

The type of data available for the two catalogues is identical, and after retrieving data \citep{ciza1,ciza2} from VizieR, we merged them into a single CIZA catalogue where the names {\tt CIZAI} and {\tt CIZAII} define the sub-catalogues (see Table \ref{table:nrobj}).

\subsection{Serendipitous catalogues}

We compiled data from a further seven serendipitous surveys as described below.

\subsubsection{160SD}

The {\bf 160SD} (The 160 Square Degree ROSAT Survey, \citealt{160sd}) is based on the serendipitous detection of extended X-ray emission in 647 archival ROSAT PSPC observations. With the galactic plane ($\mid b \mid \le 30 \, {\rm deg}$) and the regions of the Magellanic clouds excluded, the resulting sky coverage at high fluxes is 160 deg$^2$. 

A  wavelet algorithm is used to detect galaxy clusters and the quoted total luminosities are computed from the detected fluxes by assuming a $\beta$-profile with $\beta=2/3$ and a fitted core radius.  

We retrieved the full dataset \citep{160sd} from VizieR and selected only sources identified as galaxy clusters.

\subsubsection{400SD}

The {\bf 400SD} (The 400 Square Degree ROSAT PSPC Galaxy Cluster Survey, \citealt{400sd}) extends the 160SD methodology to additional PSPC observations by adopting less restrictive selection criteria (e.g., galactic latitude and absorption, exposure times). A total of 1610 fields, corresponding to a total survey area of 397 deg$^2$, are analysed to yield a large flux-limited  ($0.5-2$ keV band flux $ \ge 1.4 \times 10^{-13} \, {\rm erg} \, {\rm s}^{-1} \, {\rm cm}^{-2} $) cluster catalogue. 400SD data is available for serendipitously and not entirely serendipitously detected clusters (clusters at redshift very close to the target redshift). 

We retrieved data \citep{400sdC} from VizieR, and merged the information into a unique 400SD catalogue, introducing the sub-catalogue labels {\tt 400SD\_SER} and {\tt 400SD\_NONSER} to distinguish between the two classes of objects (see Table \ref{table:nrobj}). 

%
\begin{table}
\caption{Number of clusters in the catalogues used to construct the MCXC before and after handling of multiple entries.}            
\label{table:nrobj}     
\centering                          
\begin{tabular}{lrr}       
\hline
\hline
{\bf Catalogue} & Nr. of clusters & Nr. of clusters\\    
\ \ \ Sub-catalogues & Input & MCXC\\    
\hline
\\
{\bf NORAS/REFLEX} & 889  & 879 \\
\ \ \ NORAS & 445  & 437 \\
\ \ \ REFLEX & 444  &  442 \\
\\
{\bf 400SD} &  266 & 256 \\
\ \ \ 400SD$\_$SER & 242  & 236 \\
\ \ \ 400SD$\_$NONSER & 24  & 20 \\
\\
{\bf 160SD} &  199 & 90 \\
\\
{\bf BCS} &  312 & 80 \\
\ \ \ BCS & 205  & 47 \\
\ \ \ eBCS & 107  & 33 \\
\\
{\bf SGP} &  157 & 55 \\
\\
{\bf SHARC} & 69 & 30 \\
\ \ \ SHARC$\_$SOUTH & 37  & 14 \\
\ \ \ SHARC$\_$BRIGHT & 32  &  16 \\
\\
{\bf WARPS} & 159  & 78 \\
\ \ \ WARPS & 34  & 11 \\
\ \ \ WARPSII & 125 & 67 \\
\\
{\bf NEP} & 63  & 48 \\
\\
{\bf MACS} & 51  & 38 \\
\ \ \ MACS$\_$MJFV& 23  & 18 \\
\ \ \ MACS$\_$BRIGHT& 22  & 14 \\
\ \ \ MACS$\_$DIST& 6  &  6 \\
\\
{\bf CIZA} & 130  & 128 \\
\ \ \ CIZAI & 73  & 72 \\
\ \ \ CIZAII & 57  & 56 \\
\\
{\bf EMSS} & 102  & 61 \\
\ \ \ EMSS$\_$1994 & 81  & 47 \\
\ \ \ EMSS$\_$2004 & 21  & 14 \\
\\
\hline
TOTAL & 2397 & 1743 \\
\hline                                   
\end{tabular}
\end{table}

\subsubsection{SHARC Bright and SHARC Southern}

The SHARC survey is based on archival ROSAT PSPC observations. The {\bf SHARC Bright} (Bright Serendipitous High-Redshift Archival ROSAT Cluster survey, \citealt{sharcb}) is a wide area shallow survey covering a total area of 178.6 deg$^2$ with a flux limit of $1.63 \times 10^{-13} \, {\rm erg} \, {\rm s}^{-1} \, {\rm cm}^{-2}$. The {\bf SHARC Southern} (The Southern Serendipitous High-Redshift Archival ROSAT Cluster survey, \citealt{sharcs}) is a narrow area deep survey covering 17.7 deg$^2$ with a flux limit of $4.66 \times 10^{-14} \, {\rm erg} \, {\rm s}^{-1} \, {\rm cm}^{-2} $. Cluster detection is based on a wavelet and sliding-box techniques, respectively.

For both catalogues a $\beta$-profile with fixed $\beta=2/3$ and $r_{\rm c}$ is used to determine the total luminosity and a circular aperture of radius $r_{\rm 80}$, which contains 80 percent of the total flux. This implies that in addition to the extrapolated total luminosities, the aperture luminosities $L_{\rm ap} \equiv L(< r_{\rm 80})$ are available.  

Data \citep{sharcsC,sharcbC} are retrieved from VizieR and merged into a single SHARC catalogue (only sources identified as clusters are selected from \cite{sharcbC}) with sub-catalogues labelled {\tt SHARC\_SOUTH} and {\tt SHARC\_BRIGHT} (see Table \ref{table:nrobj}).

\subsubsection{WARPS and WARPSII}

The WARPS survey is also based on ROSAT PSPC observations. {\bf WARPS} (Wide Angle ROSAT Pointed Survey, \citealt{warps2002}), covers 16.2 deg$^2$ in 86 PSPC fields, while its extension {\bf WARPSII} (Wide Angle ROSAT Pointed Survey II, \citealt{warps2008}) covers  56.7 deg$^2$ in 301 PSPC fields. The WARPS survey uses the VTP algorithm for cluster detection and characterization.  

The quoted total luminosities are computed as in \cite{bcs}, but no information which allows the computation of aperture luminosities is reported.

Data \citep{warps2002C,warps2008C} are retrieved from VizieR. For both catalogues we include clusters below the nominal flux limit that defines the statistically complete sample. The two catalogues are merged into a single WARPS catalogue and {\tt WARPSI} and {\tt WARPSII} are adopted as sub-catalogue labels (see Table \ref{table:nrobj}).

\subsubsection{EMSS}

The {\bf EMSS} (Einstein Observatory Extended Medium Sensitivity Survey, \citealt{emss1990}) cluster catalogue is constructed from a flux-limited sample of sources serendipitously detected in {\it Einstein} IPC (Imaging Proportional Counter) fields at high galactic latitudes. 

Data are compiled from the tables published in \citet{emss1994} and \citet{emss2004}. While the sample presented in \citet{emss1994} is the most complete and up-to-date work on the entire EMSS cluster catalogue, \citet{emss2004} provides more reliable ASCA measurements for the $z \ge 0.3$ EMSS clusters. The {\it Einstein} luminosities reported in \citet{emss1994} are computed from the flux measured in a 2\farcm4 x 2\farcm4 detection cell by adopting a $\beta$-model with fixed $\beta=2/3$. The information provided is not sufficient to compute aperture luminosities from the quoted total luminosities. The ASCA luminosities in \citet{emss2004} are total luminosities. Since distant clusters are not resolved by ASCA, these luminosities were derived by assuming that the clusters are point sources. Hence in this case only total luminosities are available. 

Clusters in the \citet{emss2004} sample are removed from \citet[][]{emss1994}\footnote{For a comparison between EMSS and ASCA flux measurements see \citet{emss2004}.}. We remove MS1209.0+3917, MS1333.3+1725, and MS1610.4+6616 for the reasons mentioned in \citet{emss2004}. The data are then merged into a single EMSS catalogue where the names {\tt EMSS\_1994}  and {\tt EMSS\_2004} denote the sub-catalogue labels (see Table \ref{table:nrobj}).

\section{Data extraction and homogenisation}
\label{data.sec}
 
The data provided by the different input catalogues (positions, redshifts, names, luminosities, etc.) are rather similar. However some data homogenisation is needed, in particular for quantities such as luminosity and mass. 

As detailed above, in many cases the luminosity is measured within some small aperture $R_{\rm ap}$ ($L_{\rm ap} \equiv L(< R_{\rm ap})$) is the corresponding aperture luminosity) and then extrapolated to some larger radius using a reasonable model of the surface brightness profile. The radial extrapolation might be extremely large, implying that the derived luminosity is basically equal to the total luminosity $L_{\rm tot}=L(< \infty)$. Another common choice is to extrapolate to $R_{\rm 200}$. In this case the luminosity $L_{\rm 200} \equiv L(< R_{\rm 200})$ is essentially equal to $L_{\rm tot}$, since the contribution to the total luminosity of the emission between $R_{\rm 200}$ and infinity is fully negligible. With the present generation of X-ray observations, the standard choice is  $R_{\rm 500}$ ($L_{\rm 500} \equiv L(< R_{\rm 500})$), and we have chosen this radius for the MCXC data homogenisation procedure.

The assumed cosmological model is of course at the basis of our homogenisation procedure. In the following all luminosities and other cluster parameters which depend on the distance scale are converted to our reference cosmology (i.e. $\Omega_{\rm M} = 0.3$, $\Omega_\Lambda= 0.7$, and $H_{0} = 70$~ km/s/Mpc).

Below we list all the quantities that are provided by the MCXC and explain in detail how they are derived from the original information in the input catalogues. The names of the quantities as given in the associated electronic table are given in {\tt typewriter} typeface. MCXC clusters are ordered by  right ascension. As an example we list the first 40 entries by splitting the information into Tables \ref{MCXCtableBIG1} and \ref{MCXCtableBIG2}.

\subsection{Coordinates and redshifts}

The cluster coordinates given in the input catalogues are those of the cluster centroid determined from X-ray data (apart from those in the sub-catalogue {\tt EMSS\_1994} which are the coordinates of the cluster optical position). For the MCXC, all coordinates are converted to right ascension and declination for the epoch J2000 in hours (degrees), minutes, and seconds ({\tt RAJ2000} and {\tt DEJ2000}) and in units of decimal degrees ({\tt \_RAJ2000} and {\tt \_DEJ2000}). We also provide the cluster positions in galactic coordinates -- {\tt GLON} and {\tt GLAT} are galactic longitude and latitude, respectively, in degrees (see Table \ref{MCXCtableBIG1}). 

\begin{figure*}
\centering
  \includegraphics[width=0.95 \textwidth]{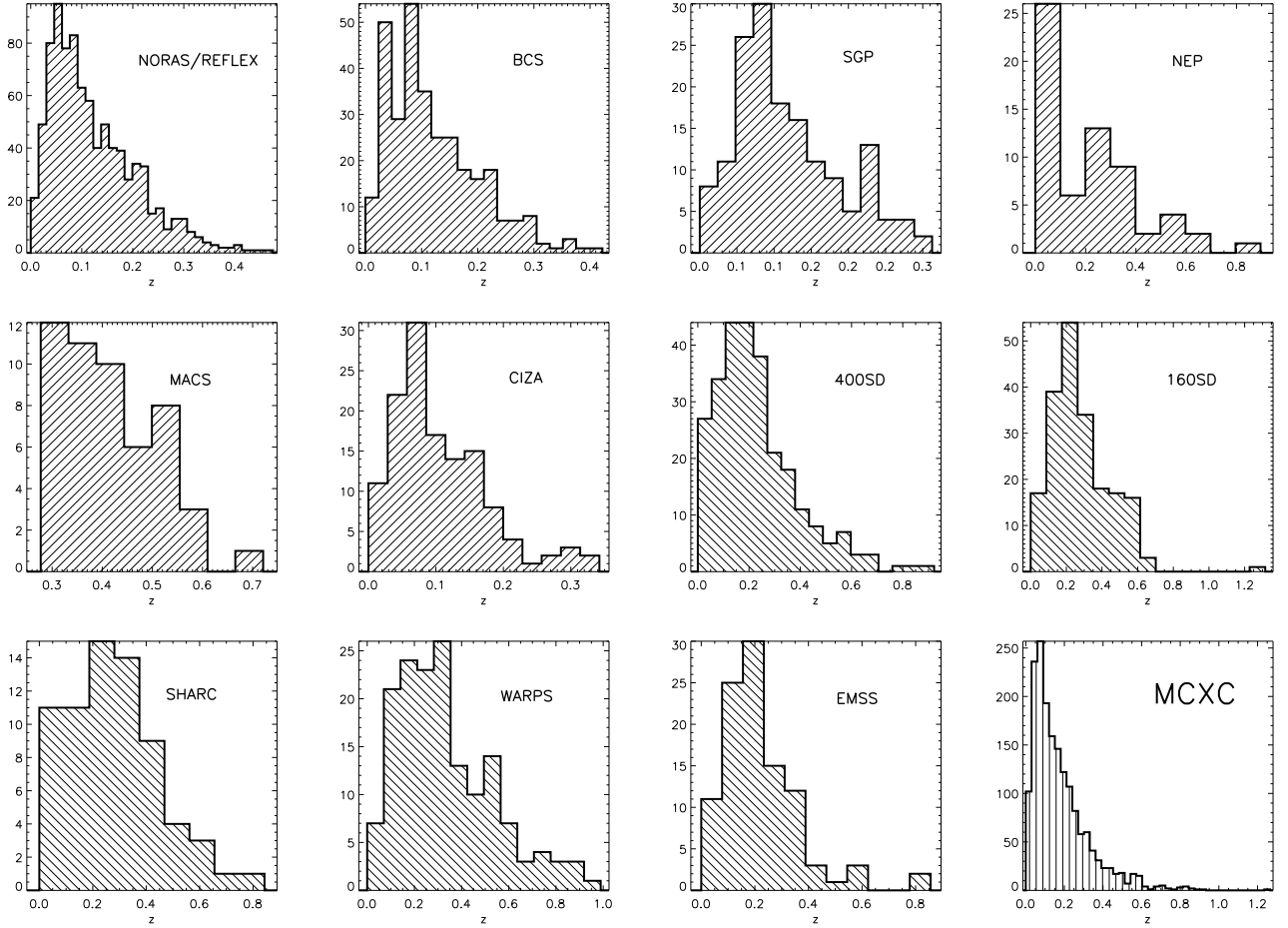}
  \caption{Redshift histograms of the input catalogues used to construct the MCXC and of the MCXC after handling of multiple entries. Different area shadings are used for RASS-based, serendipitous, and MCXC catalogues.}
  \label{zhisto:fig}
\end{figure*}

No manipulation is needed for the cluster redshifts {\tt z}. As stated above, only clusters with measured redshift are retained in the MCXC (see Table \ref{MCXCtableBIG1}). In Fig. \ref{zhisto:fig}, we show the redshift histograms of the individual input catalogues used to construct the MCXC and of the MCXC after handling of multiple entries (see Sect. \ref{overlap.sec}). The histograms highlight the different redshift ranges typically probed by serendipitous and RASS-based surveys, with the latter generally being confined to local and medium redshift clusters.    

\subsection{Names}

Two types of cluster name are usually listed in the input catalogues: the name assigned by the authors {\tt NAME}, and the alternative name {\tt NAME$\_$ALT} (see Table \ref{MCXCtableBIG1}). {\tt NAME} is usually constructed from the cluster coordinates (e.g., RXJ0041.1-2339 in 160SD, MS0007.2-3532 in EMSS, MACSJ0011.7-1523 in MACS, RXC\,J0000.1+0816 in NORAS/REFLEX, CIZA, and SGP, RX\,J1716.6+6410 in NEP and SHARC, J0022.0+0422 in WARPS). Exceptions to this format are BCS and 400SD. BCS names are listed as they appear in optical catalogues (e.g. ZwCl1432, A602), while in the 400SD, names are not assigned. We therefore assigned a {\tt NAME} to {\tt 400SD\_SER} clusters according to the standard SIMBAD\footnote{\tt http://simbad.u-strasbg.fr/simbad/} format acronym 'BVH2007 NNN' (Burenin+Vikhlinin+Hornstrup+, 2007, e.g., BVH2007 193), and for the {\tt 400SD\_NONSER} we created a new acronym 'BVH2007 NS NNN' (e.g., BVH2007 NS 12). For all but the 400SD clusters we retain the original names as listed in the input catalogues. 

Alternative names in the input catalogues are mostly based on cataloged optical counterparts to the X-ray sources (e.g. A2894, ZwCl 0104.9+5350, UGC 12890). In some cases alternative names are given as notes or comments, and we also use this information to construct {\tt NAME$\_$ALT} in the MCXC by extracting the suitable piece of information. Alternative names are homogenised so as to match both SIMBAD and NED\footnote{\tt http://nedwww.ipac.caltech.edu/} standards. When this is not possible we choose the SIMBAD acronym conventions. Moreover, when multiple alternative names are available, they are listed separated with a comma. For BCS clusters we set {\tt NAME$\_$ALT} equal to {\tt NAME}. For 160SD and 400SD clusters alternative names are extracted from the notes. For 160SD clusters the identifier 'VMF98 NNN' is also used. 

Notice that in most of the input catalogues alternative names end with letters for double systems (e.g., A2384 (A), A3574E, etc.). Such information is important because it indicates whether the measured luminosity refers to the whole system to only a part of it.  

Our choice of formats for {\tt NAME} and {\tt NAME$\_$ALT} in the MCXC is made in order to facilitate queries in the SIMBAD and NED databases. Notice that both {\tt NAME} and {\tt NAME$\_$ALT} also facilitate the handling of duplicate entries as discussed extensively below in Sect. \ref{overlap.sec}. 

In addition to the above two cluster identifiers we add a third name, {\tt NAME$\_$MCXC} (see Table \ref{MCXCtableBIG1}), that is constructed from coordinates for the epoch J2000. It allows a fully unambiguous cluster identification in the MCXC catalogue and is defined as  MCXC\,JHHMM.m+DDMM. 

\subsection{Catalogue and Sub-catalogue}

As explained above in Sect. \ref{catalogues.sec} and listed in Table \ref{table:nrobj}, for each cluster the input catalogue and sub-catalogue names are given in {\tt CATALOGUE} and {\tt SUB$\_$CATALOGUE} (see Table \ref{MCXCtableBIG1}). If no sub-catalogue exists the sub-catalogue name is equal to the catalogue name.

\begin{figure*}[]
\centering
  \includegraphics[width=1.05 \textwidth]{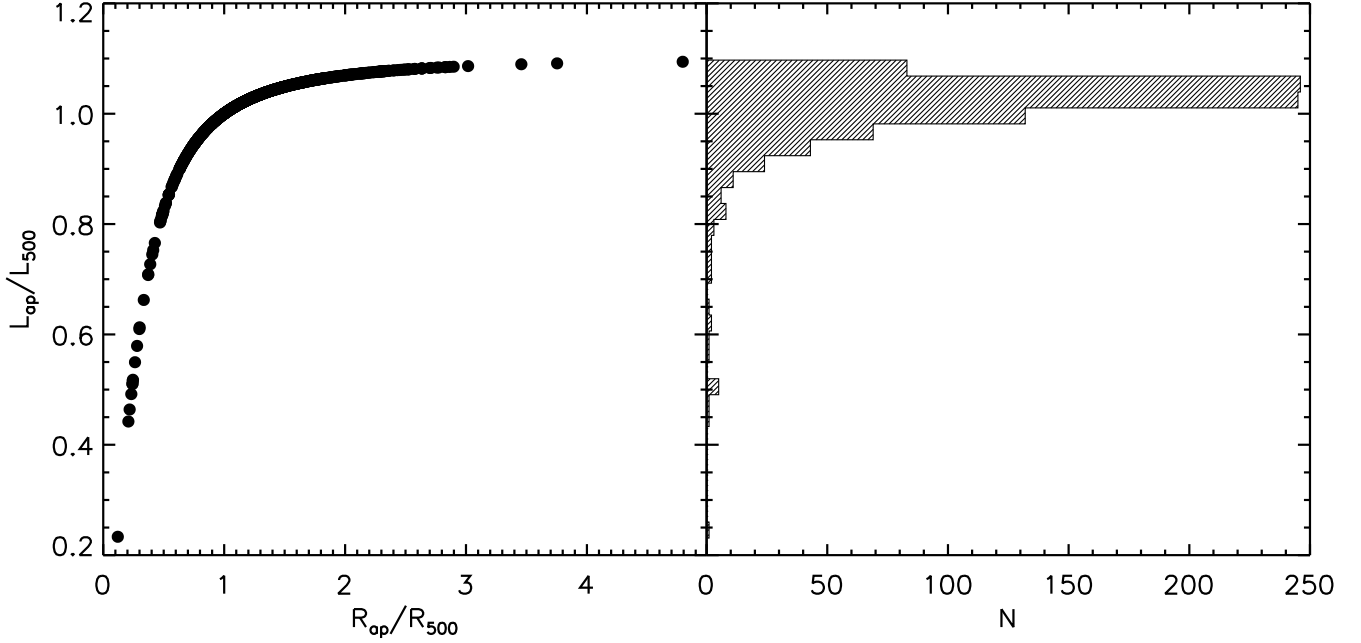}
  \caption{Relation between input quantities $R_{\rm ap}$ and $L_{\rm ap}=L(< R_{\rm ap})$ and the iteratively estimated $R_{\rm 500}$ and $L_{\rm 500}=L(< R_{\rm 500})$ for the NORAS/REFLEX clusters. The luminosity ratio as a function of aperture radius ratio is shown in the left panel, while the luminosity ratio histogram is shown in the right panel.}
     \label{apeff:fig}
\end{figure*}

\subsection{Luminosities}

The luminosities are homogenised according to the following procedure:

\begin{enumerate}

\item  When necessary, we first convert the input luminosity (e.g., $0.5-2$ keV band in {\tt NEP}, bolometric in {\tt MACS$\_$MJFV}, $0.3-3.5$ keV band in {\tt EMSS$\_$1994}) to the $0.1-2.4$ keV energy band using the MEKAL plasma code \citep{mewe1985,liedahl1995}. The temperature dependence of this conversion is taken into account by either using measured temperatures when available in the input catalogue, or by iteration about the the non-core-excised luminosity-temperature relation of \cite{pratt2009}, assuming an abundance of 0.3. In the following all the quoted luminosities are therefore as measured in the $0.1-2.4$ keV energy band for our reference cosmology.

\item The resulting luminosities are then converted to $L_{\rm 500}$ adopting two different procedures, depending on the type of luminosity measurement available in the input catalogue.

\begin{itemize}

\item If only the total luminosity $L_{\rm tot}$ (i.e., extrapolated up to large distances) is available we adopt $L_{\rm 500}=a \times L_{\rm tot}$, where $a$ is the ratio $L_{\rm 500} / L_{\rm tot}$ for a luminosity profile model based in the average gas density profile derived from the representative X-ray cluster sample \rexcess\  \citep{croston2008}. More precisely, from the individual scaled density profiles \citep[see][left panel of Figure 3]{app2009}, we computed the average profile and fitted it with the AB-model given by Eqn.~2 in \cite{pratt2002}:

\begin{equation}
\rho_{gas} \propto  \left( \frac{x}{x_c} \right)^{-\alpha} \times \left[1+ \left( \frac{x}{x_c} \right)^{2} \right]^{-3 \beta /2  + \alpha/2} ,
\label{nmean:eq} 
\end{equation}

\noindent for $x = r/R_{\rm 500}$, finding $x_{\rm c} = 0.303$, $\alpha =  0.525$, and  $\beta=  0.768$. 

Since only recently observational progress has shown that the AB-model (Eq. \ref{nmean:eq}) yields a more accurate description than the traditional $\beta$-model \citep[see][and references therein]{croston2008}, most of the analyses listed in Sect. \ref{catalogues.sec} adopted the latter when extrapolating luminosities to large radii. For the sake of clarity, in Appendix \ref{ap:beta} we illustrate the differences between luminosities computed adopting the AB-model or the $\beta$-model.    

Since observed luminosities are derived from surface brightness profiles and integration within circular apertures, the resulting luminosity profile is cylindrically integrated up to aperture radii of 1 and $5 \times R_{\rm 500}$ to compute $L_{\rm 500}$ and $L_{\rm tot}$, respectively. The cluster boundary is also assumed to be equal to $5 \times R_{\rm 500}$. We find a ratio $a=L_{\rm 500} / L_{\rm tot}=0.91$ and that the exact choice of the aperture enclosing the total luminosity is not relevant.  
This constant $\sim 10 \%$ correction is therefore applied to all the 160SD, 400SD, SGP, WARPS, CIZA, {\tt MACS$\_$DIST}, and EMSS clusters for which only total luminosities are available (see Sect. \ref{catalogues.sec}). For the {\tt MACS$\_$MJFV} and {\tt MACS$\_$BRIGHT} clusters no conversion is needed since the quoted luminosities are $L_{\rm 500}$.

\item For the remaining clusters, i.e. those with available aperture luminosities $L_{\rm ap}$ (1333 objects in total ), we compute $L_{\rm 500}$ iteratively. The basic ingredients of this iterative procedure are the luminosity profile model implied by Eq. \ref{nmean:eq} and the luminosity-mass relation (L-M relation, hereafter):

\begin{equation}
h(z)^{-7/3} \, \Big( \frac{L_{500}}{10^{44} \, erg \, s^{-1}} \Big)=C \, \Big( \frac{M_{500}}{3 \times 10^{14} M_{\odot}} \, \Big)^{\alpha},
\label{L-M:eq} 
\end{equation}

with ${\rm log}(C)=0.274$ and $\alpha=1.64$ \citep[see Table 1 in][]{app2009}. These values are slightly different from those given in \cite{pratt2009} due to \citeauthor{app2009}'s use of an updated $M_{\rm 500}-Y_{\rm X}$ relation. Specifically,  we use the relation in Eq. 2 of \cite{app2009}, i.e. we adopt a non-self-similar slope for the $M_{\rm 500}-Y_{\rm X}$ relation. The adopted $C$ and $\alpha$ values are derived from \rexcess\  luminosity data uncorrected for the Malmquist bias. The effect of these choices is further discussed below.

In addition to Eqs. \ref{nmean:eq} and \ref{L-M:eq}, our iterative procedure necessitates the basic input quantities $R_{\rm ap}$ and $L_{\rm ap}=L(< R_{\rm ap})$, the circular aperture radius and the aperture luminosity, respectively. These are either directly available from the input catalogues or can be computed as explained in Sect. \ref{catalogues.sec} for the NORAS/REFLEX, BCS, SHARC, and NEP catalogues. 

For each cluster, the aperture luminosity and radius together with the model luminosity profile set the luminosity profile in physical units (i.e. radius in Mpc). The latter are then iteratively converted in units of $R_{\rm 500}$ using Eq. \ref{L-M:eq} by using $L_{\rm ap}$ as the starting luminosity in the iteration. In Fig. \ref{apeff:fig} we illustrate the relation between the input luminosity $L_{\rm ap}$ and $L_{\rm 500}$ for the NORAS/REFLEX clusters. As expected, small/large apertures yield final luminosities $L_{\rm 500}$ which are larger/smaller that the input aperture luminosities (left panel). Notice that while on average the difference between $L_{\rm ap}$ and $L_{\rm 500}$ is $\sim 5 \%$, it is very relevant for a significant portion of the sample (right panel).

In order to explore the effect of our choice of $M_{\rm 500}-Y_{\rm X}$ relation, we iteratively estimate $L_{\rm 500}$ by adopting the Malmquist bias uncorrected L-M relation derived from the $M_{\rm 500}-Y_{\rm X}$ relation with {\it self-similar slope} given in Eq. 3 of \cite{app2009}. We find that for $\sim 96, 91 \%$ of the clusters the difference is less that $5, 2 \%$, respectively. The largest differences are found for low luminosity objects. In fact, if we consider clusters with  $L_{\rm 500} \ge 10^{43}$ erg/s) we find that for $\sim 99, 95 \%$ of the clusters the difference is less that $5, 2 \%$, respectively.

In order to explore the reliability of our assumption concerning the Malmquist bias correction of the L-M relation, we repeated our iterative procedure by using the {\it Malmquist bias corrected relation} of \citet{pratt2009}, finding essentially the same $L_{\rm 500}$ (relative differences are $\sim 1 \%$). This is expected, because the steep drop of the typical cluster luminosity profile with radius makes $L_{\rm 500}$ rather insensitive to the exact choice of $R_{\rm 500}$.

\end{itemize}

\end{enumerate}

Using the two above-described methods we can therefore systematically compute $L_{500}$, the $0.1-2.4$ keV energy band luminosities within $R_{500}$, for all the clusters (Table \ref{MCXCtableBIG2}, {\tt L$\_$500}).

\subsection{Total masses}

Total masses $M_{\rm 500}$, estimated for the same cosmology adopted here, are directly available only for {\tt MACS$\_$MJFV} and {\tt MACS$\_$BRIGHT} clusters\footnote{There are two exceptions (MACSJ0358.8-2955 and MACSJ2311.5+0338), but these clusters do not end up in the MCXC because they are also members of other catalogues.}. For almost all the clusters we therefore rely on luminosity as a mass proxy and estimate $M_{\rm 500}$ ({\tt M$\_$500} in Table \ref{MCXCtableBIG2}) using Eq. \ref{L-M:eq}. While our computation of $L_{\rm 500}$ does not depend on the details of the adopted L-M relation (non-self-similarity of the underlying $M_{\rm 500}-Y_{\rm X}$ relation and Malmquist bias correction), obviously the estimated $M_{\rm 500}$ does. In particular, the $M_{\rm 500}$ values provided by the MCXC rely on the assumption that on average the Malmquist bias for the samples used to construct the MCXC is the same as that of the \rexcess \ sample. Since the selection functions of the samples we use are complex (and indeed, in most cases are not known or available), our mass estimates must rely on this assumption. In addition, while our choice ensures maximal self-consistency in our modeling, other calibrations of the L-M relation could be adopted. Nevertheless, given our estimated $L_{\rm 500}$, the computation of total masses from a different L-M relation is straightforward.

From $M_{\rm 500}$ we estimate the characteristic radii $R_{\rm 500}$ ({\tt R$\_$500} in Table \ref{MCXCtableBIG2}) using:
\begin{equation}
M_{500}= (4 \pi/3)  \, \rho_\mathrm{c}(z) \, 500 \, R_{500}^3 \, ,
\label{r500:eq} 
\end{equation}
where the critical density is $\rho_\mathrm{c}(z)=3 H(z)^{2} /8 \pi G $.

\subsection{Notes}

We gather together useful information concerning individual objects and add it to the MCXC as notes ({\tt NOTES} in Table \ref{MCXCtableBIG2}). In the input catalogues this information is usually provided as notes and comments and because it is different in type and size from catalogue to catalogue its homogenisation is not straightforward. In general, we choose not to include detailed and extended information and we therefore refer the reader to the cited papers for more information, e.g., the notes to Table 1 in \citet{emss1994} or in Table 2 of \cite{sharcb}. In the following we describe the type of information we included in {\tt NOTES}. For the meaning of abbreviations we refer the reader to the cited papers of each sub-catalogue.

For BCS, SGP, SHARC, NEP, MACS, CIZA, and EMSS no information is provided or is too detailed to be added in concisely. For NORAS we take information from column ID in Table~1 of \citet{noras} (information on source identification). For REFLEX, \citep{reflex}, where the provided information is fairly detailed, we merge the following: (i) information in column Cm of Table 1 (information on source identification), (ii) the information concerning groupings as given in Table 10 with the simple note GR1, GR2, \ldots, GR10 if the cluster is listed in one of the 10 groupings listed in the table, (iii) multipeak information as given in Table 11 (columns Morphology and Orientation are merged, as e.g., two maxima/NE-SW), and (iv)  information on whether the cluster is part of a line of sight structure as given in Table 12 (we simply add losStr if the cluster appears in the table). For 400SD clusters we take the information given in the column Notes in Table~4 of \citet[][information on alternative names is not used]{400sd}. For 160SD clusters we take the information given in the column Notes in Table~4 of \citet[][information on alternative names is not used]{160sd}. 

\subsection{Scale}

In order to facilitate the conversion between angular and physical sizes (e.g., for $R_{\rm 500}$) we provide the angular scale factor {\tt SCALE} in kpc/arcsec (see Table \ref{MCXCtableBIG2}).

\section{Duplicate clusters}
\label{overlap.sec}

The overlap between the survey areas of the input catalogues induces duplicate (and in some cases triplicate, quadruplicate, etc.) entries in the MCXC catalogue. In this Section we explain how these are identified and which entry is retained in the MCXC. In short, we search where a given cluster is a member of each input catalogue and, according to criteria based on the type of data and the size of the input catalogue, we retain only one entry in the MCXC. The full list of clusters without removal of multiple entries can be requested from the authors.

The most important criterion that we use to decide which duplicate cluster is preserved in the MCXC is the size of the input catalogue. In addition we give higher priority to catalogues that provide aperture luminosities because they ensure the most reliable and self-consistent computation of $L_{\rm 500}$. These two criteria allow us to rank the input catalogues from highest to lowest priority as in Table \ref{table:nrobj}. Obviously, because of its size and the type of information it provides, NORAS/REFLEX is the catalogue with the highest priority. It is followed by other large and well-defined catalogues such as the 400SD, 160SD, BCS, etc. Hence, when a cluster is listed in more than one catalogue it is retained only as an entry in the input catalogue with higher priority. This catalogue ranking is not crucial for CIZA because the overlap of its survey area with other surveys is minimal.

Our procedure therefore reduces to the identification of multiple entries. This identification is based mainly on centroid coordinate differences, and to a lesser extent on redshift differences. Given the large number of entries, cluster identification is performed in three steps in order to progressively reduce the number of candidate multiple entries. 

\begin{enumerate}

\item In a first step if two clusters in different catalogues have centroid offsets of less than 1' and their relative redshift difference is less than 10 per cent they are identified as being the same cluster. Although this step removes a large number of duplicate entries, we compare their names and alternative names to make sure that we do not remove single entries. In the case of a doubtful association we do not remove any entry. 

\item In a second step the resulting catalogue is inspected once more by carefully identifying clusters with centroid offsets of less than 5, 10, or even 20', and by varying the relative redshift difference. This time consuming procedure is needed because different source analysis techniques can yield rather different centroid positions, in particular for nearby clusters. Redshift differences can be very large and we use them only as indicators and not as stringent constraints. Each multiple entry candidate is checked, with names and alternative names used to facilitate the procedure. Again in this step we are rather conservative and do not remove any cluster if the identification is not certain. 

\item In a third step the cleaned catalogue is inspected once more with large allowances for centroid offsets, and any overlaps are checked by visually inspecting RASS and PSPC maps. The associations inspected in this last step are either multiple systems or entries where very different redshifts are given for the same X-ray cluster. 

\end{enumerate}

In each of the three steps we make some exceptions to the general rules explained above. When the redshift difference is very large we keep the cluster with more recent or reliable redshift measurement. An extreme example is MCXC J1524.6+0957 (VMF98 170 or BVH2007 198), whose redshift is 0.07800 in SHARC and 0.5160 in 400SD. In this case, the average temperature of 5.1 keV measured by \cite{vik2002} rules out the low redshift estimate. For double or multiple systems we retain measurements for each of the components instead of measurements of the whole system. This explains why, although it has the highest priority, the NORAS/REFLEX catalogue finally contains 10 clusters less than before our handling of duplicates. If possible, we compared our duplicates identifications with those in other work \citep[e.g.,][]{160sd} and find perfect agreement.

In order to retain useful information, when an entry with no available alternative name is kept in the MCXC catalogue while the one we discard provides it, we copy this information into the retained entry.  

The MCXC provides information concerning multiple entries in the input catalogues though the label {\tt CAT$\_$OV} (see Table \ref{MCXCtableBIG2}) which contains the name of the sub-catalogue from which the removed cluster entry is a member.  

\section{Discussion}
\label{discussion.sec}

\subsection{Global catalogue characteristics}

The final MCXC is constructed from the input catalogues discussed in Sect. \ref{catalogues.sec} with information homogenised as explained in Sect. \ref{data.sec}. Multiple entries in the resulting catalogue are handled as described in Sect. \ref{overlap.sec}. This procedure yields the final MCXC catalogue, which comprises in total 1743 clusters (2397 initially, see Table \ref{table:nrobj}) and contains virtually no multiple entries. In the following we illustrate some basic properties of the MCXC. Notice that, because of the priorities we assign to the input catalogues, NORAS/REFLEX clusters constitute a large fraction of the MCXC (see Table \ref{table:nrobj}).     

\begin{figure}[h]
\centering
  \includegraphics[width=0.5 \textwidth]{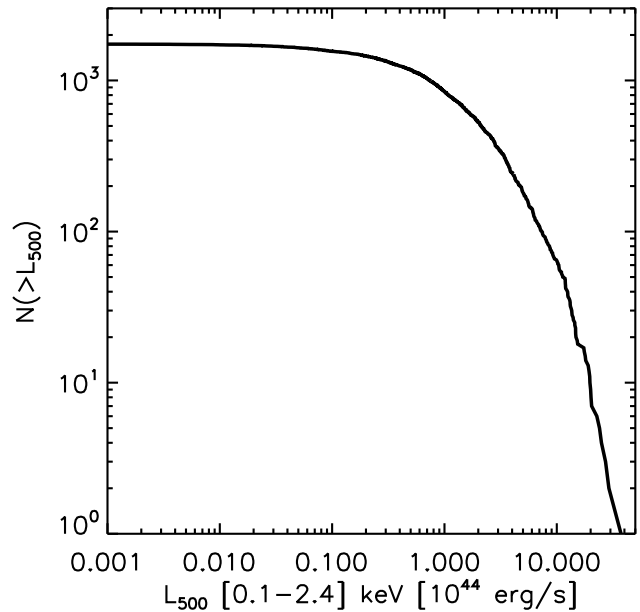}
  \caption{Luminosity distribution of MCXC clusters.}
  \label{zcumdistr:fig}
\end{figure}

\begin{figure*}
\centering
  \includegraphics[width=1.0 \textwidth]{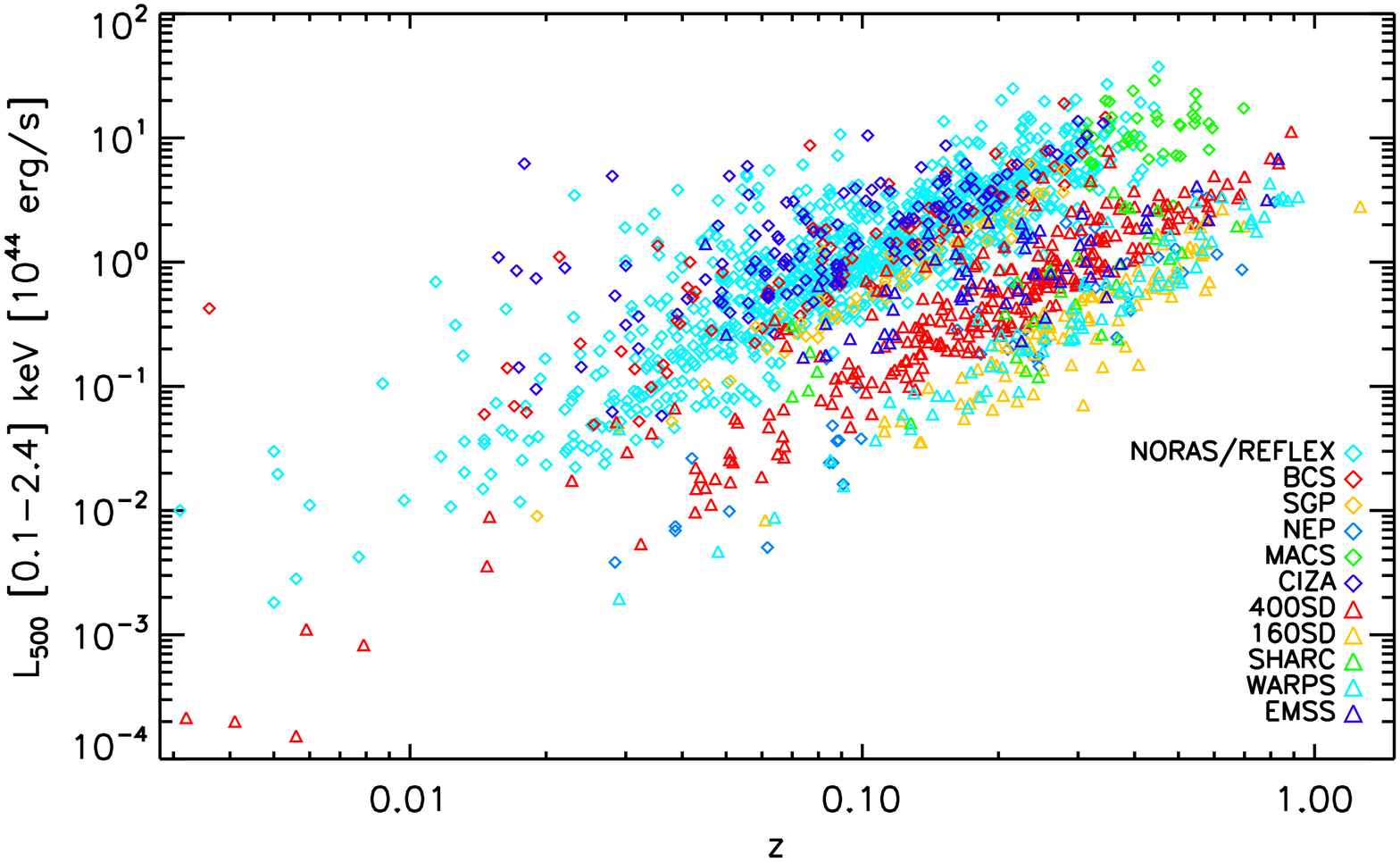}
  \includegraphics[width=1.0 \textwidth]{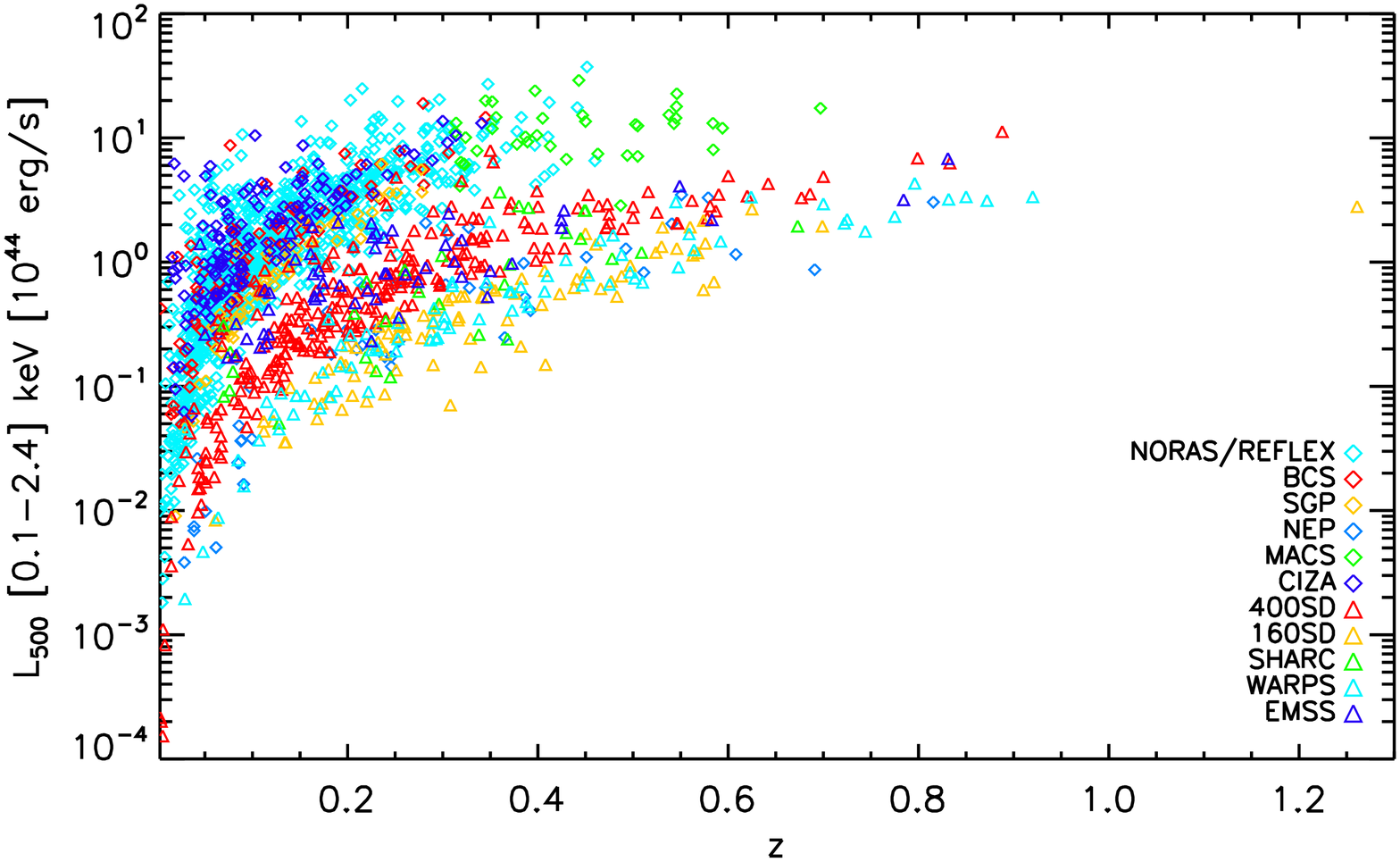}

  \caption{{\it Top}: The $0.1-2.4$ keV band luminosities $L_{\rm 500}$ of the 1743 MCXC clusters as a function of redshift. Diamonds and triangles indicate clusters from RASS-based and serendipitous  catalogues, respectively. {\it Bottom}: Same, but in lin-log scale.}
    \label{Lzloglog:fig}
\end{figure*}

The MCXC redshift histogram is illustrated in the bottom right hand panel of Fig. \ref{zhisto:fig}: 282, 77, and 18 clusters ($\sim 16, 4, 1$ per cent, respectively) have redshifts larger than 0.3, 0.5 and 0.7. In Fig~\ref{zcumdistr:fig} we show the number of clusters as a function of luminosity: 846, 64 ($\sim 49, 4$ per cent, respectively) of the clusters have $0.1-2.4$ keV band luminosities $L_{\rm 500}$ larger than $1, 10 \times 10^{44}$ erg/s.
%
%
In Fig. \ref{Lzloglog:fig} we show the $0.1-2.4$ keV band luminosities $L_{\rm 500}$ of the 1743 MCXC clusters as a function of redshift in log-log (top panel) and the more conventional lin-log scale (bottom panel). These figures highlight both the different nature of RASS-based and serendipitous surveys and their complementarity. For a given redshift, serendipitously discovered clusters are less luminous than those from RASS-based catalogues because the deeper exposures allow lower flux limits to be adopted. This implies that the fraction of high redshift clusters in serendipitous surveys is much higher than that of RASS-based surveys.    
%

\begin{figure*}
\centering
  \includegraphics[width=1. \textwidth]{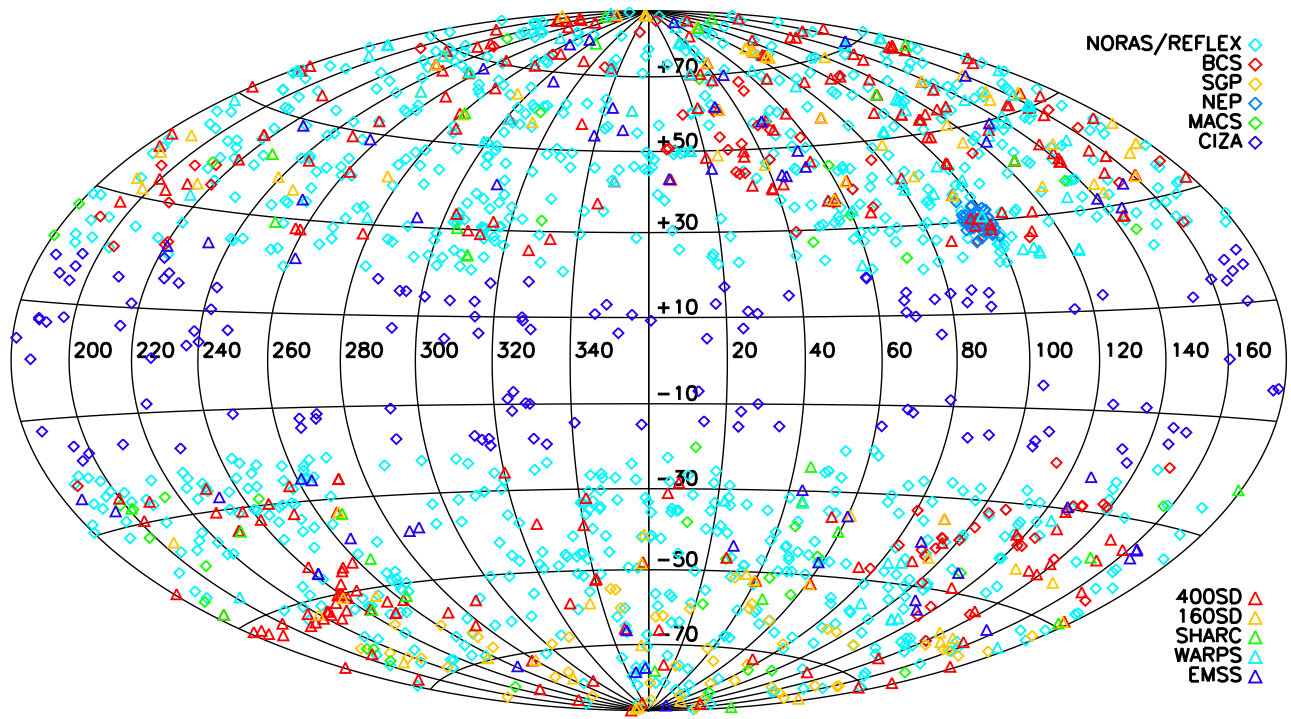}
  \caption{Sky map of the 1743 MCXC clusters in galactic coordinates. Symbols and colors are the same as in Fig. \ref{Lzloglog:fig}.}
        \label{skymap:fig}
\end{figure*}

In addition to redshift and luminosity (and total mass), a fundamental quantity provided by the MCXC is the cluster position in the sky, both in equatorial and galactic coordinates. In Fig. \ref{skymap:fig} we show the distribution on the sky of the 1743 MCXC clusters in galactic coordinates. Some distinctive features are: NORAS/REFLEX, BCS and MACS clusters are fairly homogeneously distributed; the only clusters at low galactic latitude are from the CIZA survey; the RASS-based clusters of SGP and NEP are localised in narrow regions; serendipitous clusters are sparsely distributed across the sky.
%

\subsection{Robustness of luminosity measurements}

As the luminosity $L_{\rm 500}$ the most relevant physical quantity provided by the MCXC, we focus on its discussion in the remainder of this Section.

\begin{figure}
\centering
  \includegraphics[width=0.5 \textwidth]{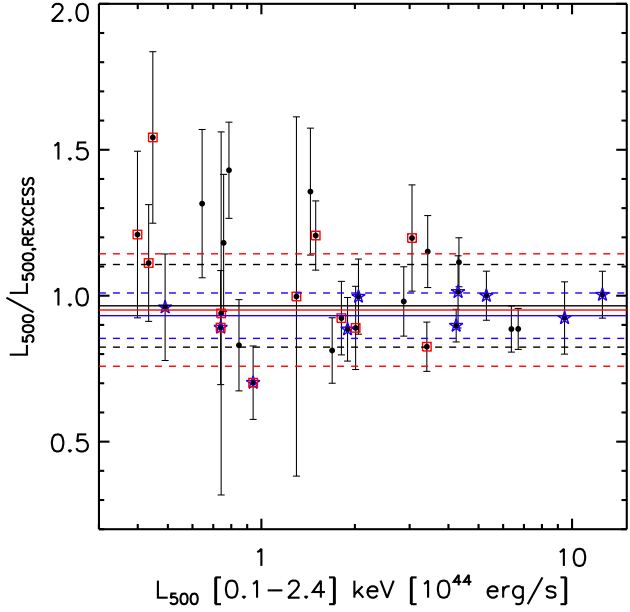}
  \caption{Ratio $L_{\rm 500}/L_{\rm 500,\rexcess}$ between our estimate of the $0.1-2.4$ keV band luminosities $L_{\rm 500}$ and the {\it XMM-Newton} measurements of \cite{pratt2009} as a function of $L_{\rm 500}$.  Blue stars indicate cool core clusters while red squares morphologically disturbed clusters. Solid lines indicate error weighted means and dashed lines represent the error weighted means $\pm$ error weighted standard deviations (black for all clusters, blue for cool core clusters, and red for morphologically disturbed clusters as defined in \cite{pratt2009}).}
  \label{rexcesscomp}
\end{figure}

Since the modeling adopted in Sect. \ref{data.sec} is based on results from the \rexcess \ sample, and that the latter is a subsample of REFLEX, the comparison of the $L_{\rm 500}$ values derived in this work and those given in \cite{pratt2009} for the 31 \rexcess\  clusters provides a useful test for our procedure. We remind the reader that for all the REFLEX clusters we computed the luminosities $L_{\rm 500}$ from aperture luminosities by means of the iterative procedure explained in Sect. \ref{data.sec}. We find that our derived $R_{\rm 500}$ is larger than $R_{\rm ap}$ for only seven \rexcess\  clusters, and at most only by $\sim 20$ per cent. In Fig. \ref{rexcesscomp} we show the ratio between our estimate of $L_{\rm 500}$ and the {\it XMM-Newton} measurements given in \cite{pratt2009}, $L_{\rm 500,\rexcess}$. Uncertainties on the luminosity ratios are computed from quadratic sum of the errors given in \cite{pratt2009} and propagation of the aperture luminosity errors provided in \cite{reflex}. Notice that for one cluster the redshift adopted in our work differs from the one used in $L_{\rm 500,\rexcess}$. Although we correct for this difference, this has no impact on our results. Symbols in Fig. \ref{rexcesscomp} are as in \cite{pratt2009}, i.e. blue stars for cool core clusters and red squares for morphologically disturbed clusters. We compute error weighted means and standard deviations of the luminosity ratio and find: $0.965 \pm 0.141$ for all 31 clusters, $0.932 \pm 0.078$ for the cool core clusters, and $0.951 \pm 0.193$ for the disturbed clusters. Our comparison indicates a good agreement between the two measurements. Notice the fairly large scatter at low luminosity, the large scatter for disturbed clusters with respect to cool core clusters, and that there is an indication that our luminosity estimates are on average biased low in cool core systems (a $1 \, \sigma$ effect). The lower luminosity ratio for cool core clusters is expected because they are modeled using the AB-model derived from the mean of the \rexcess \ sample (Eq. \ref{nmean:eq}), although their emission is more  centrally peaked. We find no trend of luminosity ratio with the ratio $R_{\rm ap}/R_{\rm 500}$ and no redshift dependence as the \rexcess \ sample redshift leverage is too small. 
%
%

\subsection{Intercomparison of original luminosity measurements}

The procedure adopted to handle multiple entries (detailed in Sect. \ref{overlap.sec} ) allows us to compare $L_{\rm 500}$ estimates derived from different input luminosity measurements. A total of 558 MCXC entries list the properties of a cluster that is a member of more that one input catalogue, and for which only the information from only one input catalogue has been retained. For these entries there are therefore N luminosity measurements (one provided by the MCXC and N-1 unused overlaps) of the $0.1-2.4$ keV band luminosity $L_{\rm 500}$. There are 5 clusters with $N=5$, 8 with $N=4$, 59 with $N=3$, and 486 with $N=2$. 

The luminosities $L_{\rm 500}$ are compared as follows. For each of the 558 entries we compute the ratio $L_{\rm 500}/L_{\rm 500, MCXC}$ where $L_{\rm 500, MCXC}$ is the luminosity given in the MCXC and $L_{\rm 500}$ is the luminosity of the same cluster, but derived from a different input catalogue (i.e., the overlap luminosity). As explained in Sect. \ref{overlap.sec}, in some cases the redshifts provided by the input catalogues can be fairly different. We therefore correct the luminosities of the overlaps by multiplying them with the squared ratio of the luminosity distances at the two different redshifts. This is equivalent to comparing the $0.1-2.4$ keV band fluxes within $R_{\rm 500}$. The MCXC provides these luminosity ratios though the quantity {\tt L$\_$500$\_$RAT} (see Table \ref{MCXCtableBIG2}) where they are ordered in the same way as the sub-catalugue names in {\tt CAT$\_$OV}. 

\begin{figure}
  \resizebox{ \textwidth}{!}
{\includegraphics[bb=0 300 1097 1097]{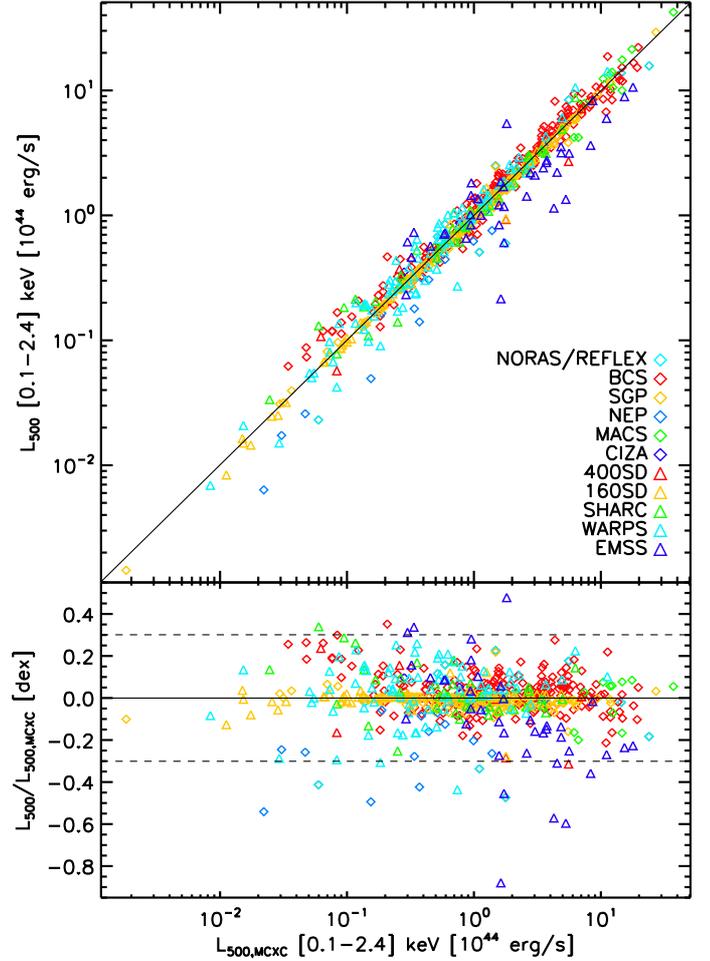}}
  \caption{Luminosity $L_{\rm 500}$ of the overlaps (top panel) and the ratio $L_{\rm 500}/L_{\rm 500, MCXC}$ in dex (bottom panel) as a function of $L_{\rm 500, MCXC}$, the luminosity measurements retained in the MCXC. Symbols and colors are the same as in Fig. \ref{Lzloglog:fig}, but refer to the overlaps only. The solid line in the top panel indicates the best fit to the data. The solid line in the bottom panel indicates the error weighted mean of the luminosity ratio, while the dashed horizontal lines indicates the luminosity ratios equal to 2 and 0.5.}
  \label{L500ratios}
\end{figure}

In Fig. \ref{L500ratios} we show the luminosity $L_{\rm 500}$ of the overlaps (top panel) and the ratio $L_{\rm 500}/L_{\rm 500, MCXC}$ (bottom panel, in dex units) as a function of $L_{\rm 500, MCXC}$. Both a direct fit to the data (which is shown in the top panel of the figure and basically indistinguishable from equality) and the mean value of the luminosity ratios indicate that the different luminosity determinations are in excellent agreement. In particular, in order to avoid any bias introduced by the wide range of luminosity values, we compute the ratios of the log values of the luminosities. The error weighted mean and standard deviation of these ratios are adopted to quantify the agreement between the different luminosity estimates. Errors are computed from the uncertainties quoted in the input catalogues, assuming that the relative error on the luminosities $L_{\rm 500}$ is the same as the one on the input luminosities. We find  0.999 and 0.003 for the error weighted mean and standard deviation, respectively. We performed the same analysis by taking into account whether the compared luminosities are derived from RASS or pointed observations and whether they are computed iteratively or just by adopting a constant conversion factor (see Sect. \ref{data.sec}) and find no signifiant trend. The luminosity comparison therefore shows that on average the agreement between different $L_{\rm 500}$ measurements is excellent. However, the clear outliers in Fig. \ref{L500ratios} indicate that for some clusters there are large discrepancies. 

Although a discussion on individual objects, and thus on the difference between specific survey measurements, is beyond the scope of our work, we briefly discuss very discrepant luminosity estimates by focussing on strong outliers with luminosity ratios larger than 2 or smaller 0.5 in Fig. \ref{L500ratios} (i.e., differences larger than a factor of 2). There is a total of twenty objects ($\sim 4$ per cent of those with more than one luminosity measurement) of such discrepant estimates. For six of these clusters, three luminosity estimates are available. Interestingly, of these, we always find that only one of the three is very different from the others, and that the two remaining estimates agree within a few percent. For the other fourteen clusters only two estimates of $L_{\rm 500}$ are available. Of these, five involve measurements from the EMSS and seven are faint objects at low redshift, where extrapolation might strongly affect the luminosity estimates. For the remaining two clusters (A2507 and RXC J1003.0+3254) we find no obvious explanation.
%
%
%
\section{Summary and conclusions}
\label{conclusions.sec}

Motivated by the strong need for a large, homogeneous compilation in the framework of X-ray, SZ and other multi-wavelength studies, we have presented the construction and properties of the MCXC, a Meta-Catalogue of X-ray detected Clusters of galaxies. The MCXC is constructed from publicly-available RASS-based (NORAS, REFLEX, BCS, SGP, NEP, MACS, and CIZA) and serendipitous (160SD, 400SD, SHARC, WARPS, and EMSS) cluster catalogues (see Sect. \ref{catalogues.sec}). The information from these input catalogues is systematically homogenised using the most up to date knowledge of the structural properties and scaling relations of X-ray clusters, and is undertaken in a self-consistent way (see Sect. \ref{data.sec}). More specifically, in addition to the fairly straightforward standardisation of quantities such as coordinates and redshifts ({\tt RAJ2000}, {\tt DEJ2000}, etc., and {\tt Z} in Tables \ref{MCXCtableBIG1} and \ref{MCXCtableBIG2}), we converted the available luminosities to $0.1-2.4$ keV band luminosities $L_{\rm 500}$ ( {\tt L$\_$500} in Table \ref{MCXCtableBIG2}) by adopting the average gas density profile \citep{croston2008} and L-M relation \citep{pratt2009} derived from the representative X-ray cluster sample \rexcess. The computation is performed directly from aperture luminosities when available ($\sim 76$ per cent of the MCXC clusters) and we verify that the derived luminosities do not depend on the details of the adopted L-M relation. 

Total masses $M_{\rm 500}$ and radii $R_{\rm 500}$ can be computed from the luminosities $L_{\rm 500}$ by adopting an L-M relation. The MCXC provides these quantities computed self-consistently using the L-M relation adopted in this work ({\tt M$\_$500} and {\tt R$\_$500} in Table \ref{MCXCtableBIG2}). The MCXC further provides three cluster identifiers: the MCXC name, the original name as given in the input catalogues, and an alternative name ({\tt NAME$\_$MCXC}, {\tt NAME}, and {\tt NAME$\_$ALT} in Table \ref{MCXCtableBIG1}, respectively). The latter has been homogenised to match both SIMBAD and NED standards. In addition, we collated important information usually provided as notes and comments in the input catalogues ({\tt NOTES } in Table \ref{MCXCtableBIG2}).

Multiple entries originating from overlaps between the survey areas of the input catalogues are very carefully handled (see Sect. \ref{overlap.sec}). The result of this procedure is provided by the MCXC ({\tt CAT$\_$OV} in Table \ref{MCXCtableBIG2}). We compare luminosity measurements from different catalogues, finding that on average the agreement is excellent,  and discuss the most discrepant measurements (see Sect. \ref{discussion.sec}). We find good agreement with the precisely measured  \rexcess\  luminosities given in \cite{pratt2009}. These comparisons strongly support the validity of our approach. The MCXC provides the luminosity ratios for clusters that appear in multiple input catalogues (see {\tt L$\_$500$\_$RAT}  in Table \ref{MCXCtableBIG2}).  

The MCXC comprises 1743 clusters ordered by  right ascension, and contains virtually no multiple entries. The full MCXC is available at CDS\footnote{\tt http://cds.u-strasbg.fr/} and contains the information given in Columns (1)-(19) in Tables \ref{MCXCtableBIG1} and \ref{MCXCtableBIG2}, where the first 40 entries are given as an example. 

We envisage that the catalogue will be useful for the construction of representative samples for work on structural properties and samples for cosmological investigations; examination of SZ-X-ray scaling relations; checking of SZ candidates; selection function studies; definition of samples for lensing and optical/IR follow-up. The work by \cite{melin10} is an example of how the information provided by the MCXC can be used for SZ studies. In particular, their work illustrates that given the MCXC luminosities, the universal pressure profile and the associated SZ scaling relations provided by \cite{app2009} yield the X-ray predicted SZ signal from individual objects, which can be then compared to those observed with WMAP. This approach will be extremely useful for studies based on SZ surveys such as Planck, SPT, and ACT.

The MCXC is an ongoing project and will be extended to include available data for individual clusters at high redshift (the most relevant for cosmological studies) and cluster catalogues derived from ongoing X-ray surveys when they are publicly available.

\begin{acknowledgements}
We thank Isabella Gioia, Patrick Henry, and Donald Horner for helpful discussions and Nabila Aghanim for strongly supporting the project. This research has made use of the VizieR database, operated at CDS, Strasbourg. EP acknowledges the support of grant ANR-06-JCJC-0141.

\end{acknowledgements}


\begin{landscape}
\begin{table}
\caption{The first 40 entries of the MCXC catalogue. The full MCXC is available at CDS.}\label{MCXCtableBIG1}
\centering
{\tiny
\begin{tabular}{lllrrrrrrcll}
\hline\hline            
{\tt NAME$\_$MCXC } & {\tt NAME} &{\tt NAME$\_$ALT } & {\tt RAJ2000 }& {\tt DEJ2000 }& {\tt \_RAJ2000 } & {\tt \_DEJ2000 } & {\tt GLON } & {\tt GLAT } & {\tt Z } &  {\tt CATALOGUE } & {\tt SUB$\_$CATALOGUE }\\
\hline
MCXC name & Name & Alternative & Right & Declination & Right & Declination  & Galactic & Galactic & Redshift & Catalogue & Sub-catalogue\\
 &  & name & ascension& & ascension & &  longitude & latitude & & name & name \\
 &  &  &(J2000)&(J2000) & (deg) & (deg) & (deg) & (deg) &  & & \\
\hline
(1)&(2)&(3)&(4)&(5)&(6)&(7)&(8)&(9)&(10)&(11)&(12) \\
\hline
MCXC J0000.1+0816 & RXC J0000.1+0816 & UGC 12890 & 00 00 07.1 & 08 16 27.8 & 0.030 & 8.274 & 101.783 & -52.477 & 0.0396 & NORAS/REFLEX & NORAS \\
MCXC J0000.4-0237 & RXC J0000.4-0237 &  & 00 00 24.7 & -02 37 30.0 & 0.103 & -2.625 & 94.268 & -62.622 & 0.0379 & SGP & SGP \\
MCXC J0001.6-1540 & RXC J0001.6-1540 &  & 00 01 39.0 & -15 40 52.0 & 0.413 & -15.681 & 75.129 & -73.733 & 0.1246 & SGP & SGP \\
MCXC J0001.9+1204 & RXC J0001.9+1204 & A2692 & 00 01 57.0 & 12 04 22.8 & 0.488 & 12.073 & 104.308 & -49.001 & 0.2033 & NORAS/REFLEX & NORAS \\
MCXC J0003.1-0605 & J0003.1-0605 & A2697 & 00 03 11.8 & -06 05 09.6 & 0.799 & -6.086 & 92.169 & -66.033 & 0.2320 & NORAS/REFLEX & REFLEX \\
MCXC J0003.2-3555 & J0003.2-3555 & A2717 & 00 03 12.1 & -35 55 37.6 & 0.801 & -35.927 & 349.330 & -76.490 & 0.0490 & NORAS/REFLEX & REFLEX \\
MCXC J0003.8+0203 & J0003.8+0203 & A2700 & 00 03 50.6 & 02 03 48.2 & 0.961 & 2.063 & 99.610 & -58.637 & 0.0924 & NORAS/REFLEX & REFLEX \\
MCXC J0004.9+1142 & RXC J0004.9+1142 & UGC 00032 & 00 04 59.4 & 11 42 02.2 & 1.247 & 11.701 & 105.239 & -49.569 & 0.0761 & NORAS/REFLEX & NORAS \\
MCXC J0005.3+1612 & RXC J0005.3+1612 & A2703 & 00 05 22.6 & 16 12 37.8 & 1.344 & 16.211 & 107.133 & -45.244 & 0.1164 & NORAS/REFLEX & NORAS \\
MCXC J0006.0-3443 & J0006.0-3443 & A2721 & 00 06 03.0 & -34 43 26.8 & 1.513 & -34.724 & 352.147 & -77.668 & 0.1147 & NORAS/REFLEX & REFLEX \\
MCXC J0006.3+1052 & RXC J0006.3+1052 & ZwCl15 & 00 06 21.7 & 10 52 03.7 & 1.591 & 10.868 & 105.386 & -50.462 & 0.1698 & NORAS/REFLEX & NORAS \\
MCXC J0008.9+4110 & ZwCl28 & ZwCl28 & 00 08 56.9 & 41 10 37.2 & 2.237 & 41.177 & 114.386 & -20.989 & 0.1537 & BCS & eBCS \\
MCXC J0009.7-3516 & MS0007.2-3532 &   & 00 09 46.5 & -35 16 30.0 & 2.444 & -35.275 & 347.884 & -77.942 & 0.0500 & EMSS & EMSS$\_$1994 \\
MCXC J0011.3-2851 & J0011.3-2851 & A2734 & 00 11 20.7 & -28 51 18.4 & 2.836 & -28.855 & 19.562 & -80.986 & 0.0620 & NORAS/REFLEX & REFLEX \\
MCXC J0011.7-1523 & MACSJ0011.7-1523 &  & 00 11 42.8 & -15 23 22.0 & 2.928 & -15.389 & 82.746 & -75.067 & 0.3780 & MACS & MACS$\_$BRIGHT \\
MCXC J0011.7+3225 & RXC J0011.7+3225 & A0007 & 00 11 44.4 & 32 25 01.2 & 2.935 & 32.417 & 113.289 & -29.710 & 0.1073 & NORAS/REFLEX & NORAS \\
MCXC J0013.6-1930 & J0013.6-1930 & A0013 & 00 13 38.3 & -19 30 07.6 & 3.409 & -19.502 & 72.276 & -78.456 & 0.0940 & NORAS/REFLEX & REFLEX \\
MCXC J0014.3-6604 & J0014.3-6604 & A2746 & 00 14 18.4 & -66 04 39.0 & 3.577 & -66.078 & 308.850 & -50.622 & 0.1599 & NORAS/REFLEX & REFLEX \\
MCXC J0014.3-3023 & J0014.3-3023 & A2744 & 00 14 18.8 & -30 23 00.2 & 3.578 & -30.383 & 8.936 & -81.240 & 0.3066 & NORAS/REFLEX & REFLEX \\
MCXC J0014.3+0854 & MS0011.7+0837 &   & 00 14 19.8 & 08 54 00.0 & 3.583 & 8.900 & 107.635 & -52.866 & 0.1630 & EMSS & EMSS$\_$1994 \\
MCXC J0015.4-2350 & J0015.4-2350 & A14 & 00 15 24.0 & -23 50 42.0 & 3.850 & -23.845 & 52.930 & -81.233 & 0.0645 & NORAS/REFLEX & REFLEX \\
MCXC J0015.9+1614 & MS0013.4+1558 &   & 00 15 55.9 & 16 14 57.5 & 3.983 & 16.249 & 110.669 & -45.775 & 0.0830 & EMSS & EMSS$\_$1994 \\
MCXC J0016.3-3121 & RXC J0016.3-3121 & A2751 & 00 16 19.8 & -31 21 55.1 & 4.083 & -31.365 & 1.882 & -81.252 & 0.0805 & SGP & SGP \\
MCXC J0016.7+0646 & RXC J0016.7+0646 & A0016 & 00 16 45.5 & 06 46 25.0 & 4.190 & 6.774 & 107.775 & -55.074 & 0.0833 & NORAS/REFLEX & NORAS \\
MCXC J0017.5-3509 & J0017.5-3509 & A2755 & 00 17 33.7 & -35 09 54.0 & 4.390 & -35.165 & 342.856 & -79.187 & 0.0968 & NORAS/REFLEX & REFLEX \\
MCXC J0018.5+1626 & MACSJ0018.5+1626 & CL 0016+1609 & 00 18 33.8 & 16 26 16.6 & 4.641 & 16.438 & 111.609 & -45.710 & 0.5456 & MACS & MACS$\_$DIST \\
MCXC J0019.0-2026 & RXC J0019.0-2026 & S26 & 00 19 03.9 & -20 26 17.2 & 4.766 & -20.438 & 73.324 & -80.025 & 0.2773 & SGP & SGP \\
MCXC J0019.6+2517 & RXC J0019.6+2517 &  & 00 19 39.1 & 25 17 26.9 & 4.913 & 25.291 & 113.924 & -37.024 & 0.1353 & NORAS/REFLEX & NORAS \\
MCXC J0020.1+0005 & RXC J0020.1+0005 &  & 00 20 10.7 & 00 05 30.1 & 5.044 & 0.092 & 106.230 & -61.762 & 0.2124 & NORAS/REFLEX & NORAS \\
MCXC J0020.5-4913 & RXC J0020.5-4913 & A2764 & 00 20 34.1 & -49 13 40.1 & 5.142 & -49.228 & 315.963 & -67.112 & 0.0711 & SGP & SGP \\
MCXC J0020.6+2840 & RXC J0020.6+2840 & A0021 & 00 20 40.9 & 28 40 30.4 & 5.171 & 28.675 & 114.819 & -33.712 & 0.0940 & NORAS/REFLEX & NORAS \\
MCXC J0020.7-2542 & J0020.7-2542 & A0022 & 00 20 42.8 & -25 42 37.1 & 5.179 & -25.710 & 42.851 & -82.978 & 0.1410 & NORAS/REFLEX & REFLEX \\
MCXC J0021.5+2803 & RXC J0021.5+2803 & IV Zw 015 & 00 21 35.9 & 28 03 04.7 & 5.400 & 28.051 & 114.954 & -34.358 & 0.0948 & NORAS/REFLEX & NORAS \\
MCXC J0022.0+0422 & J0022.0+0422 & GHO 00190.5+0405 & 00 22 03.4 & 04 22 37.9 & 5.514 & 4.377 & 109.128 & -57.705 & 0.4070 & WARPS & WARPS \\
MCXC J0023.1+0421 & J0023.1+0421 &  & 00 23 06.0 & 04 21 13.0 & 5.775 & 4.354 & 109.596 & -57.783 & 0.4530 & WARPS & WARPS \\
MCXC J0024.0-1704 & RXC J0024.0-1704 & A2768 & 00 24 03.6 & -17 04 32.2 & 6.015 & -17.076 & 89.329 & -78.121 & 0.1890 & SGP & SGP \\
MCXC J0024.5+3312 & RXC J0024.5+3312 &  & 00 24 31.8 & 33 12 31.3 & 6.133 & 33.209 & 116.478 & -29.326 & 0.2260 & NORAS/REFLEX & NORAS \\
MCXC J0025.4-1222 & MACSJ0025.4-1222 &  & 00 25 29.4 & -12 22 37.1 & 6.372 & -12.377 & 99.264 & -74.044 & 0.5843 & MACS & MACS$\_$DIST \\
MCXC J0025.5-3302 & J0025.5-3302 & S0041 & 00 25 32.4 & -33 02 49.9 & 6.385 & -33.047 & 344.774 & -81.854 & 0.0491 & NORAS/REFLEX & REFLEX \\
MCXC J0026.7+0501 & J0026.7+0501 & GHO 0024+0444 & 00 26 47.8 & 05 01 25.7 & 6.699 & 5.024 & 111.519 & -57.301 & 0.2529 & WARPS & WARPSII \\
 \hline
\end{tabular}
}
\end{table}
\end{landscape}

\begin{landscape}
\begin{table}
\caption{The first 40 entries of the MCXC catalogue, continued.}\label{MCXCtableBIG2}
\centering
{\tiny
\begin{tabular}{lllllllll}
\hline\hline            
{\tt NAME$\_$MCXC } & {\tt Z } & {\tt SCALE } & {\tt L$\_$500 } & {\tt M$\_$500 } & {\tt R$\_$500 } & {\tt NOTES } & {\tt CAT$\_$OV } & {\tt L$\_$500$\_$rat} \\
\hline
MCXC name & Redshift & Scale & $L_{\mathrm{500}}$ & $M_{\mathrm{500}}$ & $R_{\mathrm{500}}$ & Notes  & Catalogues & $ L_{\mathrm{500}} / L_{\mathrm{500,MCXC}}$ \\
 & & (kpc/") & ($10^{44}$ erg/s) & ($10^{14} \, M_{\odot}$) & (Mpc) &   & overlap & \\
 \hline
 & &(13)&(14)&(15)&(16)&(17)&(18)&(19) \\
\hline
MCXC J0000.1+0816 & 0.0396 & 0.784 & 0.196 & 0.737 & 0.630 &  & BCS & 1.084 \\
MCXC J0000.4-0237 & 0.0379 & 0.752 & 0.052 & 0.330 & 0.482 &   &  &   \\
MCXC J0001.6-1540 & 0.1246 & 2.234 & 0.815 & 1.656 & 0.802 &   &  &   \\
MCXC J0001.9+1204 & 0.2033 & 3.342 & 1.990 & 2.693 & 0.918 &  &  &   \\
MCXC J0003.1-0605 & 0.2320 & 3.698 & 6.107 & 5.219 & 1.133 &   & SGP & 0.952 \\
MCXC J0003.2-3555 & 0.0490 & 0.959 & 0.442 & 1.202 & 0.739 & losStr & SGP & 0.886 \\
MCXC J0003.8+0203 & 0.0924 & 1.719 & 0.847 & 1.734 & 0.823 &   & eBCS$\, \vert \,$SGP & 0.920$\, \vert \,$0.922 \\
MCXC J0004.9+1142 & 0.0761 & 1.443 & 0.519 & 1.301 & 0.752 &  & eBCS & 0.967 \\
MCXC J0005.3+1612 & 0.1164 & 2.107 & 1.579 & 2.493 & 0.922 & B & EMSS$\_$1994 & 0.533 \\
MCXC J0006.0-3443 & 0.1147 & 2.080 & 1.809 & 2.712 & 0.949 &   & SGP & 0.949 \\
MCXC J0006.3+1052 & 0.1698 & 2.895 & 2.273 & 2.994 & 0.962 &  & eBCS & 0.933 \\
MCXC J0008.9+4110 & 0.1537 & 2.668 & 2.111 & 2.896 & 0.957 &   &  &   \\
MCXC J0009.7-3516 & 0.0500 & 0.977 & 0.262 & 0.873 & 0.664 &   &  &   \\
MCXC J0011.3-2851 & 0.0620 & 1.195 & 1.086 & 2.061 & 0.881 & losStr & SGP & 0.914 \\
MCXC J0011.7-1523 & 0.3780 & 5.188 & 8.900 & 7.200 & 1.190 &   &  &   \\
MCXC J0011.7+3225 & 0.1073 & 1.962 & 2.572 & 3.378 & 1.023 &  & BCS & 1.042 \\
MCXC J0013.6-1930 & 0.0940 & 1.746 & 1.236 & 2.182 & 0.888 & losStr & SGP & 0.955 \\
MCXC J0014.3-6604 & 0.1599 & 2.756 & 2.827 & 3.446 & 1.012 & X &  &   \\
MCXC J0014.3-3023 & 0.3066 & 4.522 & 11.818 & 7.361 & 1.236 &   & SGP$\, \vert \,$MACS$\_$BRIGHT & 0.985$\, \vert \,$1.139 \\
MCXC J0014.3+0854 & 0.1630 & 2.800 & 1.928 & 2.722 & 0.934 &   &  &   \\
MCXC J0015.4-2350 & 0.0645 & 1.240 & 0.326 & 0.988 & 0.689 & X & SGP & 0.956 \\
MCXC J0015.9+1614 & 0.0830 & 1.561 & 0.320 & 0.964 & 0.679 &   &  &   \\
MCXC J0016.3-3121 & 0.0805 & 1.518 & 0.495 & 1.261 & 0.743 &   &  &   \\
MCXC J0016.7+0646 & 0.0833 & 1.566 & 0.319 & 0.963 & 0.679 &  & eBCS & 1.385 \\
MCXC J0017.5-3509 & 0.0968 & 1.792 & 0.692 & 1.529 & 0.788 & losStr & SGP & 0.964 \\
MCXC J0018.5+1626 & 0.5456 & 6.386 & 17.911 & 7.785 & 1.148 &   & EMSS$\_$2004 & 0.593 \\
MCXC J0019.0-2026 & 0.2773 & 4.215 & 5.571 & 4.763 & 1.081 &   &  &   \\
MCXC J0019.6+2517 & 0.1353 & 2.397 & 1.442 & 2.327 & 0.895 &  &  &   \\
MCXC J0020.1+0005 & 0.2124 & 3.458 & 0.687 & 1.398 & 0.735 &  &  &   \\
MCXC J0020.5-4913 & 0.0711 & 1.356 & 0.268 & 0.873 & 0.659 &   &  &   \\
MCXC J0020.6+2840 & 0.0940 & 1.746 & 1.435 & 2.389 & 0.916 &  & BCS & 1.201 \\
MCXC J0020.7-2542 & 0.1410 & 2.482 & 2.872 & 3.527 & 1.026 &   & SGP & 0.912 \\
MCXC J0021.5+2803 & 0.0948 & 1.759 & 0.968 & 1.878 & 0.845 &  & BCS & 0.956 \\
MCXC J0022.0+0422 & 0.4070 & 5.430 & 0.582 & 1.082 & 0.628 &  &  &   \\
MCXC J0023.1+0421 & 0.4530 & 5.781 & 0.785 & 1.250 & 0.647 &  &  &   \\
MCXC J0024.0-1704 & 0.1890 & 3.156 & 1.484 & 2.276 & 0.872 &   &  &   \\
MCXC J0024.5+3312 & 0.2260 & 3.626 & 2.993 & 3.394 & 0.983 &  &  &   \\
MCXC J0025.4-1222 & 0.5843 & 6.603 & 8.042 & 4.623 & 0.950 &   &  &   \\
MCXC J0025.5-3302 & 0.0491 & 0.961 & 0.495 & 1.287 & 0.756 &   & SGP & 0.929 \\
MCXC J0026.7+0501 & 0.2529 & 3.943 & 0.326 & 0.860 & 0.616 &  &  &   \\
 \hline
\end{tabular}
}
\end{table}
\end{landscape}

\begin{appendix} 
\section{AB-model versus $\beta$-model luminosity profiles}
\label{ap:beta}

In the following we illustrate the difference between predictions based on the AB-model and a 'typical' $\beta$-model. In particular we focus on $L_{\rm 500}$ and total luminosities $L_{\rm tot}$ estimated from a given aperture luminosity.

The AB-model adopted in this work is given by Eq. \ref{nmean:eq} with $x = r/R_{\rm 500}$, $x_{\rm c} = 0.303$, 
$\alpha =  0.525$, and  $\beta=  0.768$ (see Sect. \ref{data.sec}) and we investigate $\beta$-models given by Eq. \ref{nmean:eq} with 
$\alpha =  0$, $\beta= 2/3$, and $x_{\rm c} = 0.05, 0.1,0.2,0.3, \, {\rm and} \,  0.4$ ($x_{\rm c} = r_{\rm c}/R_{\rm 500}$, where $r_{\rm c}$ is 
the usual $\beta$-model core radius).  

\begin{figure}
  \includegraphics[width=0.5 \textwidth]{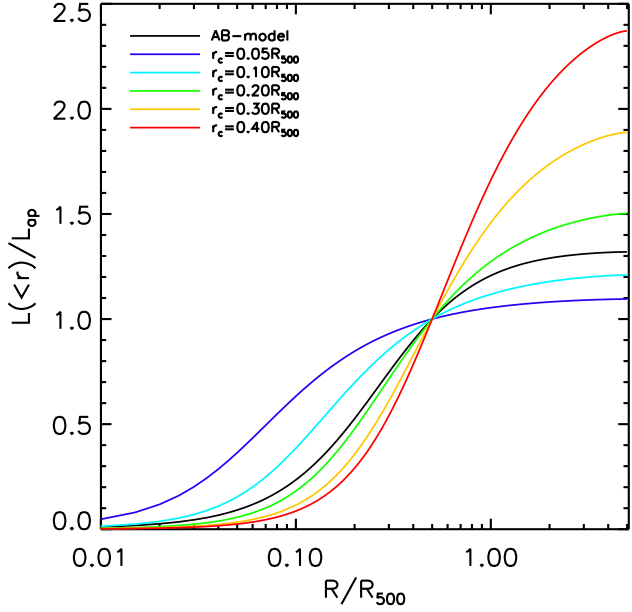}
  \caption{Luminosity radial profile normalised at $R_{\rm ap}=0.5 \times R_{\rm 500}$ for the AB-model (black line) and $\beta$-models with different core radii (color lines).}
  \label{ABvsbeta1:fig}
\end{figure}

For all models we compute luminosity profiles (spherically symmetric) which are then cylindrically integrated to obtain 'projected' luminosities 
as a function of cluster-centric distance. Finally these are normalised at $R_{\rm ap}$ where $L(< R_{\rm ap})=L_{\rm ap}$ and shown 
in Fig. \ref{ABvsbeta1:fig} for $R_{\rm ap}=0.5 \times R_{\rm 500}$. The figure shows that, with respect to the AB-model, $\beta$-models with small core radii yield centrally concentrated luminosity profiles, which in turn are shallow at large radii. The opposite is true for $\beta$-models with large core radii, which predict very extended profiles. With respect to the AB-model, a $\beta$-model with $x_{\rm c} = 0.05$ underestimates $L_{\rm 500}$ by $\sim 20 \%$ while for $x_{\rm c} = 0.4$ it yields a factor of 2 larger value.

\begin{figure}
  \includegraphics[width=0.5 \textwidth]{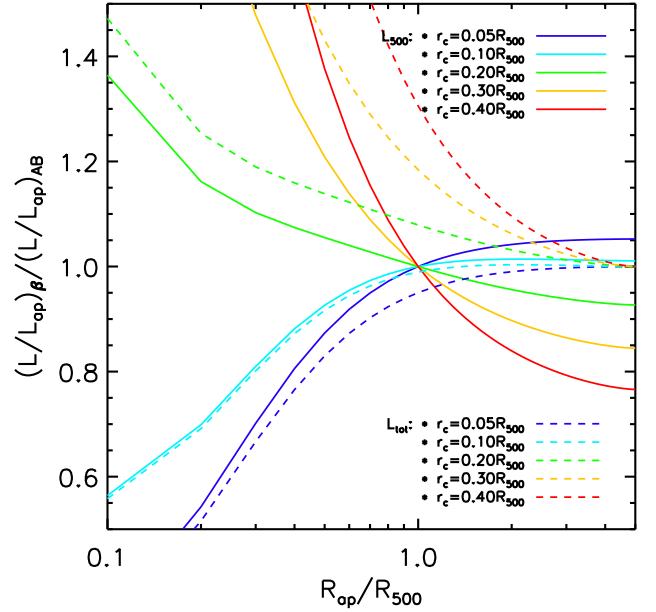}
  \caption{The $\beta$-model to AB-model ratio of the normalised luminosity profile $L/L_{\rm ap}$ evaluated at $R_{\rm 500}$ (solid lines) and $5 \times R_{\rm 500}$ (dashed lines) as a function of $R_{\rm ap}$. }
  \label{ABvsbeta2:fig}
\end{figure}

We investigate the effect of modeling on global luminosities by computing $L_{\rm 500}$ and $L_{\rm tot}=L(< 5 \times R_{\rm 500})$ as a function of $R_{\rm ap}$. In Fig. \ref{ABvsbeta2:fig} we show the $\beta$-model to AB-model ratio of $L/L_{\rm ap}$ (the normalised luminosity profile) evaluated at $R_{\rm 500}$ (i.e. $L=L_{\rm 500}$, solid lines) and $5 \times R_{\rm 500}$ (i.e. $L=L_{\rm tot}$, dashed lines) as a function of $R_{\rm ap}$. With respect to the AB-model, $\beta$-models with small/large core radii underestimate/overestimate $L_{\rm 500}$ for apertures smaller that $R_{\rm 500}$, while for $R_{\rm ap} > R_{\rm 500}$ this behavior is reversed. Total luminosities $L_{\rm tot}$ are always higher/lower than the AB-model estimates for $\beta$-models with small/large core radii and the difference increases with decreasing $R_{\rm ap}$.

%
%
%
%

\end{appendix}



\begin{thebibliography}{}
   
\bibitem[Andersson et al.(2010)]{andersson2010} Andersson,
K., Benson, B.~A., Ade, P.~A.~R., et al.\ 2010, ApJ, submitted ({\tt 
arXiv:1006.3068})

\bibitem[Arnaud et 
al.(2007)]{arnaud2007} Arnaud, M., Pointecouteau, E., \& Pratt, G.~W.\ 2007, \aap, 474, L37

\bibitem[Arnaud et al.(2010)]{app2009} Arnaud, M., Pratt,
G.~W., Piffaretti, R., et al.\ 2009, A\&A, in press ({\tt arXiv:
0910.1234})

\bibitem[Barkhouse et al.(2006)]{barkhouse2006} Barkhouse, W.~A., 
Green, P.~J., Vikhlinin, A., et al.\ 2006, \apj, 645, 955

\bibitem[B{\"o}hringer et al.(1994)]{boh1994} B{\"o}hringer,
H., Briel, U.~G., Schwarz, R.~A., et al.\ 1994, \nat, 368, 828

\bibitem[B{\"o}hringer et al.(2000a)]{noras} B{\"o}hringer,
H., Voges, W., Huchra, J.~P., et al.\ 2000, \apjs, 129, 435

\bibitem[B{\"o}hringer et al.(2000b)]{norasC} B{\"o}hringer,
H., Voges, W., Huchra, J.~P., et al.\ 2000, VizieR Online Data 
Catalog, 212, 90435

\bibitem[B{\"o}hringer et al.(2004a)]{reflex} B{\"o}hringer, H., 
Schuecker, P., Guzzo, L., et al.\ 2004, \aap, 425, 367

\bibitem[B{\"o}hringer et al.(2004b)]{reflexC} B{\"o}hringer, H., 
Schuecker, P.,  Guzzo, L., et al.\ 2004, VizieR Online Data Catalog, 
342, 50367

\bibitem[B{\"o}hringer et al.(2007)]{bohrexcess} B{\"o}hringer, H., 
Schuecker, P., Pratt, G.~W., et al.\ 2007, \aap, 469, 363

\bibitem[Burenin et al.(2007)]{400sd} Burenin, R.~A.,
Vikhlinin, A., Hornstrup, A., et al.\ 2007, \apjs, 172, 561

\bibitem[Burenin et al.(2009)]{400sdC} Burenin, R.~A.,
Vikhlinin, A., Hornstrup, A., et al.\ 2009, VizieR Online Data 
Catalog, 217, 20561

\bibitem[Burke et al.(2003a)]{sharcs} Burke, D.~J., Collins,
C.~A., Sharples, R.~M., Romer, A.~K., \& Nichol, R.~C.\ 2003, \mnras, 
341, 1093

\bibitem[Burke et al.(2003b)]{sharcsC} Burke, D.~J., Collins,
C.~A., Sharples, R.~M., Romer, A.~K.,
\& Nichol, R.~C.\ 2003, VizieR Online Data Catalog, 734, 11093

\bibitem[Carlstrom et al.(2009)]{carlstrom2009} Carlstrom, J.~E., Ade, 
P.~A.~R., Aird, K.~A., et al.\ 2009, PASP, submitted ({\tt arXiv:
0907.4445})

\bibitem[Croston et al.(2008)]{croston2008} Croston, J.~H., Pratt, 
G.~W., B{\"o}hringer, H., et al.\ 2008, \aap, 487, 431

\bibitem[Cruddace et al.(2002a)]{SGP} Cruddace, R., Voges, W., 
B{\"o}hringer, H., et al.\ 2002, \apjs, 140, 239

\bibitem[Cruddace et al.(2002b)]{SGPC} Cruddace, R., Voges, W., 
B{\"o}hringer, H., et al.\ 2002, VizieR Online Data Catalog, 214, 239

\bibitem[Cruddace et al.(2003)]{SGPerr} Cruddace, R., Voges, W., 
B{\"o}hringer, H., et al.\ 2003, \apjs, 144, 299

\bibitem[Ebeling et al.(1998)]{bcs} Ebeling, H., Edge,
A.~C., Bohringer, H., et al.\ 1998, \mnras, 301, 881

\bibitem[Ebeling et al.(2000a)]{ebcs} Ebeling, H., Edge,
A.~C., Allen, S.~W., et al.\ 2000, \mnras, 318, 333

\bibitem[Ebeling et al.(2000b)]{bcsC} Ebeling, H., Edge,
A.~C., Boehringer, H., et al.\ 2000, VizieR Online Data Catalog, 730, 
10881

\bibitem[Ebeling et al.(2000c)]{ebcsC} Ebeling, H., Edge,
A.~C., Allen, S.~W., et al.\ 2000, VizieR Online Data Catalog, 731, 
80333

\bibitem[Ebeling et al.(2001)]{macs2001} Ebeling, H., Edge,
A.~C., \& Henry, J.~P.\ 2001, \apj, 553, 668

\bibitem[Ebeling et al.(2002)]{ciza1} Ebeling, H., Mullis,
C.~R., \& Tully, R.~B.\ 2002, \apj, 580, 774

\bibitem[Ebeling et al.(2007)]{ebeling2007} Ebeling, H., Barrett,
E., Donovan, D., et al.\ 2007, \apjl, 661, L33

\bibitem[Ebeling et al.(2010)]{macs2010} Ebeling, H., Edge,
A.~C., Mantz, A., et al.\ 2010, MNRAS, in press ({\tt arXiv:1004.4683})

\bibitem[Evrard et al.(1996)]{evrard1996} Evrard, A.~E., Metzler,
C.~A., \& Navarro, J.~F.\ 1996, \apj, 469, 494

\bibitem[Fassbender(2007)]{xdcp2008} Fassbender, R., 2008, PhD thesis  
({\tt astro-ph/0806.0861v1})

\bibitem[Fowler et al.(2007)]{fowler2007} Fowler, J.~W., Niemack, 
M.~D., Dicker, S.~R., et al.\ 2007, \ao, 46, 3444

\bibitem[Gioia et al.(1990)]{emss1990} Gioia, I.~M., Maccacaro,
T., Schild, R.~E., et al.\ 1990, \apjs, 72, 567

\bibitem[Gioia \& Luppino(1994)]{emss1994} Gioia, I.~M., \& Luppino, 
G.~A.\ 1994, \apjs, 94, 583

\bibitem[Henry(2000)]{henry2000} Henry, J.~P.\ 2000, \apj, 534,
565

\bibitem[Henry(2004)]{emss2004} Henry, J.~P.\ 2004, \apj, 609,
603

\bibitem[Henry et al.(2006)]{nep2006} Henry, J.~P., Mullis,
C.~R., Voges, W., et al.\ 2006, \apjs, 162, 304

\bibitem[Horner et al.(2008)]{warps2008} Horner, D.~J., Perlman,
E.~S., Ebeling, H., et al.\ 2008, \apjs, 176, 374

\bibitem[Horner et al.(2009)]{warps2008C} Horner, D.~J., Perlman,
E.~S., Ebeling, H., et al.\ 2009, VizieR Online Data Catalog, 217, 60374

\bibitem[Kocevski et al.(2007)]{ciza2} Kocevski, D.~D.,
Ebeling, H., Mullis, C.~R., \& Tully, R.~B.\ 2007, \apj, 662, 224

\bibitem[Komatsu et al.(2010)]{komatsu2010} Komatsu, E., Smith, K.~M.,
Dunkley, J., et al.\ 2010, ApJSS, submitted ({\tt arXiv:1001.4538})

\bibitem[Kravtsov et al.(2006)]{kravtsov2006} Kravtsov, A.~V.,
Vikhlinin, A., \& Nagai, D.\ 2006, \apj, 650, 128

\bibitem[Liedahl et al.(1995)]{liedahl1995} Liedahl, D.~A.,
Osterheld, A.~L., \& Goldstein, W.~H.\ 1995, \apjl, 438, L115

\bibitem[Mahdavi et al.(2008)]{mahdavi2008} Mahdavi, A., Hoekstra,
H., Babul, A., \& Henry, J.~P.\ 2008, \mnras, 384, 1567

\bibitem[Majumdar 
\& Mohr(2003)]{majumdar2003} Majumdar, S., \& Mohr, J.~J.\ 2003, \apj, 585, 603

\bibitem[Mantz et al.(2009)]{mantz1} Mantz, A., Allen, S.~W.,
Rapetti, D., \& Ebeling, H.\ 2009, MNRAS, submitted ({\tt arXiv:
0909.3098})

\bibitem[Mantz et al.(2009)]{mantz2} Mantz, A., Allen, S.~W.,
Ebeling, H., Rapetti, D., \& Drlica-Wagner, A.\ 2009, MNRAS, submitted 
({\tt arXiv:0909.3099})

\bibitem[Markevitch \& Vikhlinin(2007)]{mv07} Markevitch,~M. , \& 
Vikhlinin, A.\ 2007, \physrep, 443, 1

\bibitem[Maughan(2007)]{maughan2007} Maughan, B.~J.\ 2007, \apj, 
668, 772 

\bibitem[Maughan et al.(2008)]{maughan2008} Maughan, B.~J., Jones,
C., Forman, W., \& Van Speybroeck, L.\ 2008, \apjs, 174, 117

\bibitem[McNamara \& Nulsen(2007)]{mcnamara2007} McNamara, B.~R., \& 
Nulsen, P.~E.~J.\ 2007, \araa, 45, 117

\bibitem[Melin et al.(2006)]{melin2006} Melin, J.-B., Bartlett, J.~G., 
\& Delabrouille, J.\ 2006, \aap, 459, 341

\bibitem[Melin et al.(2010)]{melin10} Melin, J.-B., Bartlett,
J.~G., Delabrouille, J., et al.\ 2010, A\&A, submitted ({\tt arXiv:
1001.0871})

\bibitem[Menanteau et al.(2010)]{menanteau2010} Menanteau, F., 
Gonzalez, J., Juin, J.-B., et al.\ 2010, ApJ, submitted ({\tt arXiv:
1006.5126})

\bibitem[Meneghetti et al.(2010)]{meneghetti2010} Meneghetti, M., 
Rasia, E., Merten, J., et al.\ 2010, \aap, 514, A93

\bibitem[Mewe et al.(1985)]{mewe1985} Mewe, R., Gronenschild, 
E.~H.~B.~M., \& van den Oord, G.~H.~J.\ 1985, \aaps, 62, 197

\bibitem[Mullis et al.(2003)]{160sd} Mullis, C.~R., McNamara, B.~R.,
Quintana, H., et al.\ 2003, \apj, 594, 154

\bibitem[Nagai et al.(2007a)]{nagai07} Nagai, D., Vikhlinin, A.,
\& Kravtsov, A.~V. \ 2007, \apj, 655, 98

\bibitem[Pacaud et al.(2007)]{pacaud2007} Pacaud, F., Pierre, M.,
Refregier, A., et al.\ 2007, \mnras, 382, 1289

\bibitem[Perlman et al.(2002a)]{warps2002} Perlman, E.~S., Horner,
D.~J., Jones, L.~R., et al.\ 2002, \apjs, 140, 265

\bibitem[Perlman et al.(2002b)]{warps2002C} Perlman, E.~S., Horner,
D.~J., Jones, L.~R., et al.\ 2002, VizieR Online Data Catalog, 214, 265

\bibitem[Piffaretti \& Valdarnini(2008)]{pv2008} Piffaretti, R., \& 
Valdarnini, R.\ 2008, \aap, 491, 71

\bibitem[Popesso et 
al.(2004)]{popesso2004} Popesso, P., B{\"o}hringer, H., Brinkmann, J., Voges, W., \& York, D.~G.\ 2004, \aap, 423, 449 

\bibitem[Pratt \& Arnaud(2002)]{pratt2002} Pratt, G.~W., \& Arnaud, M.
\ 2002, \aap, 394, 375

\bibitem[Pratt et al.(2009)]{pratt2009} Pratt, G.~W., Croston, J.~H., 
Arnaud, M., Bohringer, H.\ 2009, \aap, 498, 361

\bibitem[Pratt et al.(2010)]{pap10}Pratt, G.~W., Arnaud, M., 
Piffaretti, R.,et al.\ 2010, \aap, 511, A85

\bibitem[Rasia et al.(2006)]{rasia06} Rasia, E., Ettori, S., 
Moscardini, L., et al. \ 2006, \mnras, 369, 2013

\bibitem[Romer et al.(2000a)]{sharcb} Romer, A.~K., Nichol, R.~C., 
Holden, B.~P., et al.\ 2000, \apjs, 126, 209

\bibitem[Romer et al.(2000b)]{sharcbC} Romer, A.~K., Nichol, R.~C., 
Holden, B.~P., et al.\ 2000, VizieR Online Data Catalog, 212, 60209

\bibitem[Romer et al.(2001)]{xcs2001} Romer, A.~K., Viana,
P.~T.~P., Liddle, A.~R., \& Mann, R.~G.\ 2001, \apj, 547, 594

\bibitem[Rosati et al.(1998)]{rdcs1998} Rosati, P., della Ceca,
R., Norman, C., \& Giacconi, R.\ 1998, \apjl, 492, L21

\bibitem[Rosati et al.(2002)]{rosati2002} Rosati, P., Borgani, S., \& 
Norman, C.\ 2002, \araa, 40, 539

\bibitem[\v{S}uhada et al.(2010)]{suh10} \v{S}uhada, R., Song, J., B
\"ohringer, H., et al. \ 2010, \aap, 514, L3

\bibitem[Sunyaev \& Zel'dovich(1972)]{sz1972} Sunyaev, R.~A., \& 
Zel'dovich, Y.~B.\ 1972, Comments on Astrophysics and Space Physics, 
4, 173

\bibitem[Tauber et al.(2010)]{tauber2010} Tauber, J.~A., Mandolesi, N., Puget, J.-L., et al.\ 2010, \aap, in press

\bibitem[Vanderlinde et al.(2010)]{vanderlinde2010} Vanderlinde, K., 
Crawford, T.~M., de Haan, T., et al.\ 2010, \apj, submitted ({\tt arXiv:
1003.0003})

\bibitem[Vikhlinin et al.(2002)]{vik2002} Vikhlinin, A., van
Speybroeck, L., Markevitch, M., Forman, W.~R., \& Grego, L.\ 2002, 
\apjl, 578, L107

\bibitem[Vikhlinin et al.(2006)]{vikh06} Vikhlinin, A., Kravtsov, A., 
Forman, W.~R. \ 2006, \apj, 640, 691

\bibitem[Vikhlinin et al.(2009)]{vik2009} Vikhlinin, A., Burenin, R.~A.,
Ebeling, H., et al.\ 2009, \apj, 692, 1033

\bibitem[Voges et al.(1999)]{rass} Voges, W., Aschenbach, B., Boller, 
T., et al.\ 1999, \aap, 349, 389

\bibitem[Voit(2005)]{voitrev2005} Voit, G.~M.\ 2005, Reviews of
Modern Physics, 77, 207

\bibitem[Zhang et al.(2010)]{zhang2010} Zhang, Y.-Y., Okabe, N.,
Finoguenov, A., et al.\ 2010, \apj, 711, 1033

\end{thebibliography}
\end{document}